\documentclass[preprint2]{aastex}
\usepackage {url} 
\def\kms{\mbox{km~s$^{-1}$}}

\def\kpch{\mbox{$h^{-1}$kpc}}

\def\LCDM{\mbox{$\Lambda$CDM}}

\def\Mpch{\mbox{$h^{-1}$ Mpc}}
\def\Mvir{\mbox{$M_{\rm vir}$}}
\def\Msub{\mbox{$M_{\rm sub}$}}

\def\Msunh{\mbox{$h^{-1}M_\odot$}}

\def\Rvir{\mbox{$R_{\rm vir}$}}
\def\Vcirc{\mbox{$V_{\rm circ}$}}

\begin{document}
\shortauthors{KLYPIN ET AL.}
\shorttitle{Halos in the standard cosmological model: results from the Bolshoi simulation.}
\title{Dark matter halos in the standard cosmological model: results from the Bolshoi simulation}

\author{Anatoly A. Klypin$^1$, Sebastian Trujillo-Gomez$^1$, and Joel Primack$^2$}

\affil{$^1$Astronomy Department, New Mexico State University,
MSC 4500, P.O.Box 30001, Las Cruces, NM, 880003-8001, USA}
\affil{$^2$Department of Physics, University of California at Santa Cruz, Santa Cruz, CA, USA}



\begin{abstract}
  Lambda Cold Dark Matter ($\Lambda$CDM) is now the standard theory of
  structure formation in the Universe.  We present the first results
  from the new Bolshoi dissipationless cosmological $\Lambda$CDM
  simulation that uses cosmological parameters favored by current
  observations.  The Bolshoi simulation was run in a volume 250
  $h^{-1}$~Mpc on a side using $\sim$8~billion particles with mass and
  force resolution adequate to follow subhalos down to the
  completeness limit of $V_{\rm circ} = 50$ km s$^{-1}$ maximum
  circular velocity.  Using merger trees derived from analysis of 180
  stored time-steps we find the circular velocities of satellites
  before they fall into their host halos. Using excellent statistics
  of halos and subhalos ($\sim$10 million at every moment and $\sim$50
  million over the whole history) we present accurate approximations
  for statistics such as the halo mass function, the concentrations
  for distinct halos and subhalos, abundance of halos as a function of
  their circular velocity, the abundance and the spatial distribution
  of subhalos. We find that at high redshifts the concentration falls
  to a minimum value of about 4.0 and then rises for higher values of
  halo mass, a new result.  We present approximations for the velocity
  and mass functions of distinct halos as a function of redshift.  We
  find that while the Sheth-Tormen approximation for the mass function
  of halos found by spherical overdensity is quite accurate at low
  redshifts, the ST formula over-predicts the abundance of halos by
  nearly an order of magnitude by $z=10$.  We find that the number of
  subhalos scales with the circular velocity of the host halo as
  $V_{\rm host}^{1/2}$, and that subhalos have nearly the same radial
  distribution as dark matter particles at radii 0.3-2 times the host
  halo virial radius.  The subhalo velocity function $N(>V_{\rm sub})$
  scales as $V_{\rm circ}^{-3}$.  Combining the results of Bolshoi and
  Via Lactea-II simulations, we find that inside the virial radius of
  halos with $\Vcirc=200~\kms$ the number of satellites is $N(>V_{\rm
    sub}) = (V_{\rm sub}/58~\kms)^{-3}$ for satellite circular
  velocities in the range $4~\kms < V_{\rm sub} < 150~\kms$.
\end{abstract}



\section{Introduction}
\label{sec:intro}
The Lambda Cold Dark Matter (\LCDM) model is the standard modern
theoretical framework for understanding the formation of structure in
the universe \citep{Dunkley09}.  With initial conditions consisting of
a nearly scale-free spectrum of Gaussian fluctuations as predicted by
cosmic inflation, and with cosmological parameters determined from
observations, \LCDM~ makes detailed predictions for the hierarchical
gravitational growth of structure. For the past several years, the
best large simulation for comparison with galaxy surveys has been the
Millennium Simulation \citep[][MS-I]{MSI}.  Here we present the first
results from a new large cosmological simulation, which we are calling
the Bolshoi simulation (``Bolshoi'' is the Russian word for
``big''\footnote{``Bolshoi'' can be translated as (1) big or large (2)
  great (3) important (4) grown-up. The Bolshoi Ballet performs in
  Bolshoi Theater in Moscow.} ).  Bolshoi has nearly an order of
magnitude better mass and force resolution than the Millennium
Run. The Millennium Run used the first-year (WMAP1) cosmological
parameters from the Wilkinson Microwave Anisotropy Probe satellite
\citep{Spergel03}.  These parameters are now known to be inconsistent
with modern measurements of the cosmological parameters.  The
Bolshoi simulation used the latest WMAP5
\citep{Hinshaw09,Komatsu09,Dunkley09} and WMAP7 \citep{Jarosik10}
parameters, which are also consistent with other recent observational
data.

The invention of Cold Dark Matter \citep{Primack84,BFPR84} soon led to
the first CDM $N$-body cosmological simulations
\citep{Melott83,DEFW85}.  Ever since then, such simulations have been
essential in order to calculate the predictions of CDM on scales where
structure has formed, since the nonlinear processes of structure
formation cannot be fully described by analytic calculations.  For
example, one of the first large simulations of the \LCDM~ cosmology
\citep{Klypin96} showed that the dark matter autocorrelation function
is much larger than the observed galaxy autocorrelation function on
scales of $\sim1$ Mpc, so ``scale-dependent anti-biasing'' was
required for \LCDM~ to match the observed distribution of galaxies.
Subsequent simulations with resolution adequate to identify the dark
matter halos that host galaxies \citep{Jenkins98,Colin99} demonstrated
that the required destruction of dark matter halos in dense regions
does indeed occur.

$N$-body simulations have been essential for determining the properties of dark matter 
halos.  It turned out that dark matter halos of all masses typically have a similar radial 
profile, which can be approximated by the NFW profile \citep{NFW96}.
Simulations were also crucial for determining the dependence of halo
concentration $c_{\rm vir} \equiv R_{\rm vir}/r_s$ and halo shape on
halo mass and redshift \citep{Bullock01,Zhao03,Allgood06,Neto07,Maccio08},
and also for determining the dependence of the concentration of halos
on their mass accretion history \citep{Wechsler02,Zhao09}. Here $R_{\rm vir}$
is the virial radius and $r_s$ is the characteristic radius where the
log-log slope of the density is equal to $-2$. Details are given in
\S3 and in Appendix B.

One of the main goals of this paper is to provide the basic statistics of
halos selected by the maximum circular velocity (\Vcirc)~ of each halo. There are
advantages to using the maximum circular velocity as compared with the
virial mass. The virial mass is a well defined quantity for distinct
halos (those that are not subhalos), but it is ambiguous for
subhalos. It strongly depends on how a particular halofinder defines
the truncation radius and removes unbound particles. It also depends on the
distance to the center of the host halo because of  tidal
stripping. Instead, the circular velocity is less prone to those
complications. Even for distinct halos the virial mass is an
inconvenient property because there are different definitions of
``virial mass''. None of them is better than the other and different
research groups prefer to use their own definition. This causes
confusion in the community and makes comparison of results less
accurate. The main motivation for using \Vcirc~ is that it is more
closely related to the properties of the central regions of halos and,
thus, to galaxies hosted by those halos.  For example, for a Milky-Way
type halo the radius of the maximum circular velocity is about 40~kpc
(and the circular velocity is nearly the same at 20~kpc), while the
virial radius is about 300~kpc. As an indication that circular
velocities are a better quantity for describing halos, we find that
most  statistics look very simple when we use circular velocities: they
are either pure power-laws (abundance of subhalos inside distinct
halos) or power-laws with nearly exponential cutoffs (abundance of
distinct halos).

\begin{deluxetable}{llllll}
\tabletypesize{\footnotesize}
\tablecolumns{6} 
\tablewidth{0pt} 
\tablecaption{Cosmological Parameters\label{tab:Cos}}
\tablehead{
           \colhead{}                                          &        
           \colhead{Hubble  $h$ }                    &               
           \colhead{$\Omega_{\rm M}$ }                 &            
           \colhead{ tilt $n$ }                            & 
           \colhead{$\sigma_8$ }  &
           \colhead{ref.}  
   }
\startdata
WMAP5  & $0.719^{+0.026}_{-0.027}$ & $0.258\pm 0.030$ & $0.963^{+0.014}_{-0.015}$ & $0.796\pm0.0326$ &\citet{Dunkley09}\\
WMAP5+BAO+SN & $0.701\pm 0.013$ & $0.279\pm 0.013$ & $0.960^{+0.014}_{-0.013}$ & $0.817\pm 0.026$ &\citet{Dunkley09}\\
X-ray clusters  & $0.715\pm 0.012$ & $0.260\pm 0.012$ & (0.95)  & $0.786\pm 0.011$ &\citet{Vikhlinin09} \\
                            &  &  & &\phantom{0.786\,}$\pm 0.020$\tablenotemark{a} & \\
X-ray clusters+WMAP5  & (0.719) & $0.30^{+0.03}_{-0.02}$ & (0.963) & $0.85^{+0.04}_{-0.02} $ &\citet{Henry09}\\
X-ray clusters+WMAP5 & $(0.72\pm 0.08)$ & $0.269 \pm 0.016$ & (0.95) & $0.82\pm 0.03$ &\citet{Mantz08}\\
\phantom{X-ray clus}+SN+BAO & & & & \\
maxBCG + WMAP5 & (0.70)                    & $0.265\pm 0.016$ & (0.96)                   & $0.807\pm 0.020$ &\citet{Rozo09}\\
WMAP7+BAO+H$_0$ & $0.704^{+0.013}_{-0.014}$ & $0.273\pm 0.014$ & $0.963\pm 0.012$ & $0.809\pm 0.024$ &\citet{Jarosik10}\\
Bolshoi simulation & 0.70 & 0.270 & 0.95 & 0.82 & --\\
Millennium simulations & 0.73 & 0.250 & 1.00 & 0.90& \citet{MSI}\\
Via Lactea-II simulation & 0.73 & 0.238 & 0.951 & 0.74& \citet{VLII}\\
\enddata
\tablecomments{Values in parentheses are priors}
\tablenotetext{a}{Systematic error}
\end{deluxetable}
\begin{figure}[htb!]
\plotone{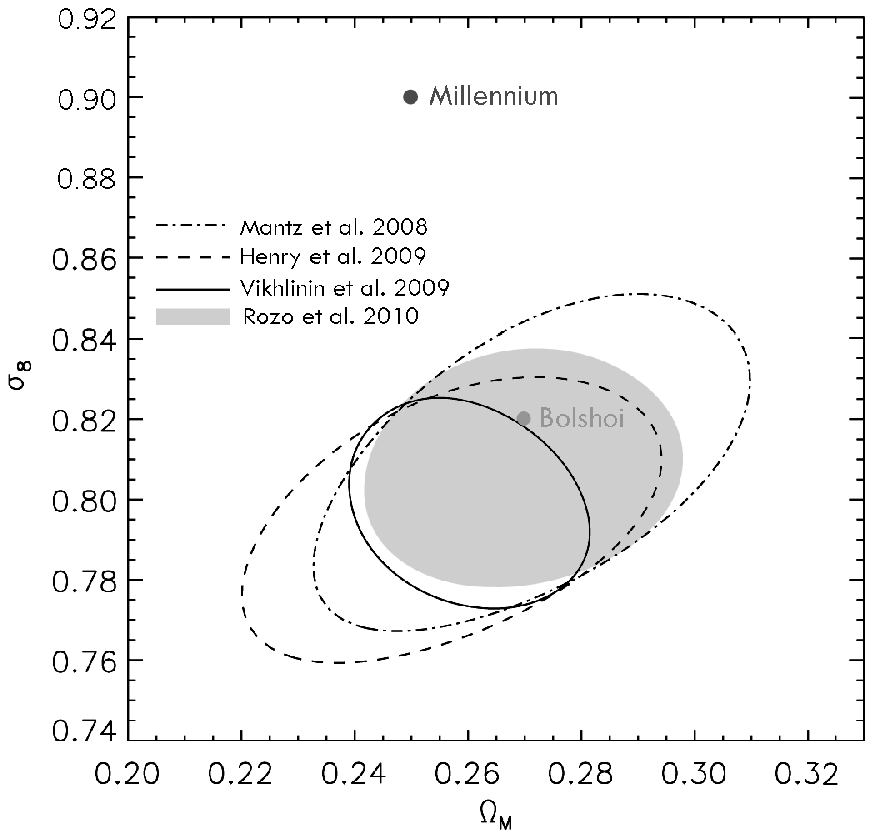}
\caption{ Optical and X-ray cluster abundance plus WMAP constraints
  on $\sigma_8$ and $\Omega_M$.  Contours show 68\% confidence regions
  for a joint WMAP5 and cluster abundance analysis assuming a flat
  $\Lambda$CDM cosmology. The shaded region is the SDSS optical maxBCG
  cluster abundance + WMAP5 analysis from \citet{Rozo09} (which is
  also the source of this figure).  The X-ray + WMAP5 constraints are
  from several sources: the low-redshift cluster luminosity function
  \citep{Mantz08}, the cluster temperature function \citep{Henry09}, and
  the cluster mass function \citep{Vikhlinin09}.  
All four recent studies
  are in excellent agreement with each other and with the Bolshoi
  cosmological parameters.  The Millennium I and II cosmological
  parameters are far outside these constraints.}
\label{fig:Cosmology}
\end{figure}

This paper is organized as follows.  In \S 2 we give the essential
features of the Bolshoi simulation.  Section~3 describes the halo
identification algorithm used in our analysis. In \S4 we present
results on masses and concentrations of distinct halos. Here we also
present relations between \Vcirc~ and \Mvir.  The halo velocity
function is presented in \S5.  Estimates of the Tully-Fisher relation
are given in \citet{Trujillo10}, where we present detailed discussions
of numerous issues related with the procedure of assigning
luminosities to halos in simulations and confront results with the
observed galaxy distribution. In \S6 we give statistics of the
abundance of subhalos. The number density profiles of subhalos are
presented in \S7. Section \S8 gives a short summary of our results.
Appendix A describes details of the halo identification procedure.
Appendix B collects useful approximation formulas. Appendix C compares
masses of halos found with two different halo-finders: the
friends-of-friends algorithm and the spherical overdensity halo-finder
used in this paper.

\section{Cosmological parameters and Simulations}
\label{sec:sims}
 The Bolshoi simulation was run with the
cosmological parameters listed in Table~\ref{tab:Cos}, together with
$\Omega_{\rm bar}=0.0469$, $n= 0.95$.  As Table~\ref{tab:Cos} shows,
these parameters are compatible with the WMAP seven-year data (WMAP7) \citep{Jarosik10}
results, with the WMAP five-year data (WMAP5), with WMAP5 combined
with Baryon Acoustic Oscillations and Supernova data \citep{Hinshaw09,
  Komatsu09,Dunkley09}. The parameters used for the Bolshoi simulation
are also compatible with the recent constraints from the Chandra X-ray
cluster cosmology project \citep{Vikhlinin09}  and
other recent X-ray cluster studies.  The Bolshoi parameters are in
excellent agreement with the SDSS maxBCG+WMAP5 cosmological parameters
\citep{Rozo09}.  The optical cluster abundance and weak gravitational lensing mass
measurements of the SDSS maxBCG cluster catalog are fully consistent
with the WMAP5 data, and the joint maxBCG+WMAP5 analysis quoted in
Table~\ref{tab:Cos} reduces the errors. Figure~\ref{fig:Cosmology}
shows current observational constraints on the $\Omega_{\rm M}$ and
$\sigma_8$ parameters.  It shows graphically that Bolshoi agrees with
the recent constraints while Millennium is far outside them.

\begin{figure}[htb!]
\plotone{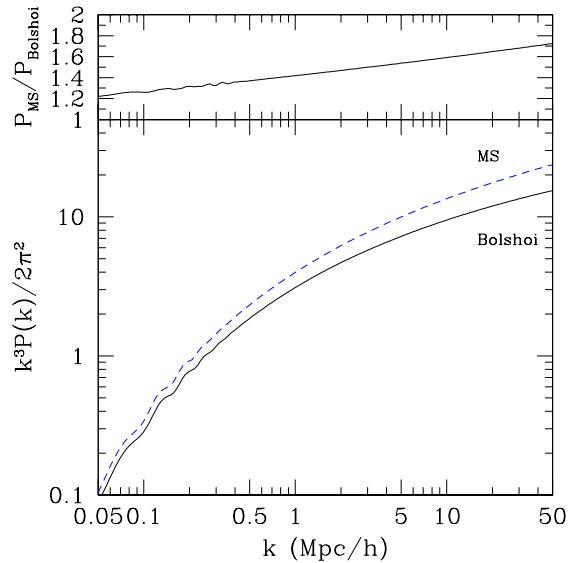}
\caption{{\it Bottom:} Linear power spectra of the Bolshoi and Millennium
  simulations at redshift zero. {\it Top:} Ratio of the spectra.  The
  Millennium simulations have substantially larger amplitude of
  perturbations on all scales, resulting in overprediction of the
  number of galaxy-size halos at high redshifts.}
\label{fig:MS}
\end{figure}

The Millennium  simulation \citep[][MS-I]{MSI} has been the basis
for many studies of the distribution and statistical properties of
dark matter halos and for semi-analytic models of the evolving galaxy
population.  However, it is important to appreciate that this
simulation and the more recent Millennium-II simulation
\citep[][MS-II]{MSII} used the first-year (WMAP1) cosmological
parameters, which are rather different from the current parameters
summarized in Table~\ref{tab:Cos}. The main difference is that the
Millennium simulations used a substantially larger amplitude of
perturbations than Bolshoi. Formally, the value of $\sigma_8$ used in
the Millennium simulations is more than 3$\sigma$ away from the
WMAP5+BAO+SN value and nearly 4$\sigma$ away from the WMAP7+BAO+H$_0$
value.  However, the difference is even larger on galaxy scales
because the Millennium simulations also used a larger tilt $n=1$ for
the power spectrum.  Figure~\ref{fig:MS} shows the linear power
spectra of the Bolshoi and Millennium simulations. Because of the large
difference in the amplitude, it is not surprising that the Millennium
simulations over-predict the abundance of galaxy-size halos. The
Sheth-Tormen approximation \citep{ST02} gives a factor of 1.3-1.7 more
$\Mvir \approx 10^{12}\Msunh$ halos at $z=2-3$ for Millennium as
compared with Bolshoi, which is a large difference. 
\citet{Angulo09b}
argue that cosmological N-body simulations can be re-scaled by certain 
approximations or by other means. However, the
accuracy of those re-scalings cannot be estimated without running
accurate simulations and testing particular characteristics.

\begin{table}[htb!]
\caption{Parameters of Bolshoi simulation\label{tab:Numb}}
\begin{tabular}{ll}
\tableline
\tableline
Parameter & Value \\
\tableline
Box size ($\Mpch$) & 250 \\
Number of particles & $2048^3$ \\
Mass resolution ($\Msunh$) & $1.35\times 10^8$ \\
Force resolution &  \\
\phantom{Maxim} ($\kpch$, physical) & 1.0 \\
Initial redshift & 80 \\
Zero-level mesh & $256^3$\\
Maximum number of  &   \\
\phantom{Maxim} refinement levels & 10  \\
Zero-level time-step $\Delta a$ & $(2-3)\times 10^{-3}$ \\
Maximum number of time-steps & $\sim 400,000$ \\
Maximum displacement  &  0.10 \\
\phantom{Maxim} per time-step &  (cell units) \\
\tableline
\end{tabular}
\end{table}

The Bolshoi simulation uses a computational box $250\Mpch$ across and
$2048^3\approx$~8 billion particles, which gives a mass resolution
(one particle mass) of $m_1=1.35\times 10^8\Msunh$. The force
resolution (smallest cell size) is physical (proper) 1~\kpch~ (see
below for details). For comparison, the Millennium-I simulation had
force resolution (Plummer softening length) 5~\kpch~ and the
Millennium-II simulation had 1~\kpch. Table~\ref{tab:Numb} gives a
short summary of various numerical parameters of the Bolshoi simulation.

The Bolshoi simulation was performed with the Adaptive-Refinement-Tree
(ART) code, which is an Adaptive-Mesh-Refinement (AMR)  type code. A detailed
description of the code is given in \citet{Kravtsov97} and \citet{Kravtsov99}.
The code was parallelized using MPI libraries and OpenMP directives
\citep{Gottlober08}. Details of the  time-stepping algorithm and comparison
with GADGET and PKDGRAV codes are given in \citet{Klypin09}. Here we
give a short outline of the code and present details specific for Bolshoi.

The ART code starts with a homogeneous mesh covering the whole cubic
computational domain. For Bolshoi we use a $256^3$ mesh.  The
Cloud-In-Cell (CIC) method is used to obtain the density on the
mesh. The Poisson equation is solved on the mesh with the FFT method
with periodic boundary conditions. The ART code increases the force
resolution by splitting individual cubic cells into $2\times 2\times
2$ cells with each new cell having  half  the size of its parent.
This is done for every cell if the density of the cell exceeds some
specified threshold. The value of the threshold varies with the level
of refinement and with the redshift. Once the hierarchy of refinement
cells is constructed, the Poisson equation is solved on each
refinement level using the Successive OverRelaxation (SOR) technique with
red-black alternations. Boundary conditions are taken from the one-level
coarser grid.  The initial guess for the gravitational potential is taken
from the previous time-step whenever possible.

\begin{figure}[tb!]
\plotone{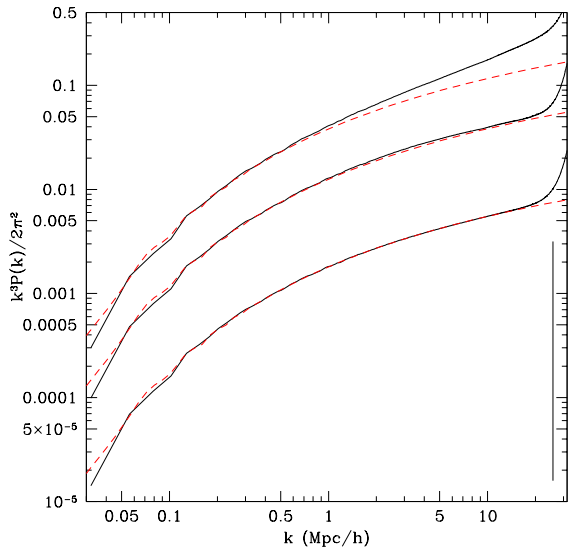}
\caption{Growth of the power spectrum of perturbations at early stages
  of evolution. The full curves show $\Delta^2 = k^3P(k)/2\pi^2$ at
  redshifts $z= 11, 20, 53$ (from top to bottom).  The dashed curves
  show the linear power spectrum. The vertical line  is the
  Nyquist frequency of particles.}
\label{fig:Pk}
\end{figure}

We use the on-line Code for Anisotropies in the Microwave Background
of \citet[CAMB][]{Lewis00}\footnote{\url{http://lambda.gsfc.nasa.gov/toolbox/tb_camb_form.cfm}}
to generate the power spectrum of cosmological perturbations. The code
is based on the CMBFAST code by \citet{CMBfast}.

At early moments of evolution, when the amplitude of perturbations was
small, the refinement thresholds were chosen in a way that allows
unimpeded growth of even the shortest perturbations, with wavelengths close
to the Nyquist frequency.  Ideally, the distance between particles should
be at least two cell sizes: at that separation the force of gravity is
Newtonian. In setting parameters for Bolshoi we came close to this
condition: we used a density threshold of 0.6 particles per cell, which
resulted in effectively resolving the whole computational volume down
to 4 levels of refinement or, equivalently, to a 4096$^3$ mesh. This condition was
kept until redshift $z=11$. As the perturbations grow, get nonlinear, and
collapse, it becomes prohibitively expensive (memory consuming) and not
necessary to keep this strict refinement condition. We gradually
increase the threshold for the 4-th refinement level: 0.8 particles
till $z=9$, 1~particle till $z=7$, 3~particles till $z=1.5$, and thereafter
we had 5~particles.  The same increase of the thresholds
was done for higher refinement levels, for which we started with
2~particles at high $z$ and ended with 5~particles at $z=0$.

Figure~\ref{fig:Pk} shows the power spectrum of perturbations at
redshifts $z=11,20, 53$ and compares it with the linear theory. A
$4096^3$ mesh was used to estimate $P(k)$ in the simulation, which may
underestimate the real spectrum by 3-5 percent at the Nyquist
frequency due to the finite smoothing of the density field produced by
the Cloud-In-Cell density assignment. Results indicate that the code
was evolving the system as it was expected to: linear growth of small
perturbations at early stages and at large scales with no suppression
at high frequencies.

The number of refinement levels, and thus the force resolution,
change with time.  For example, there were 8 levels at $z=10$, and 9
at $z=5$, which gives the proper resolution of 0.4~\kpch. At $z=1$ the
tenth level was open with the proper resolution of 0.5~\kpch. However, when a
level of refinement opens, it contains only a small fraction of volume
and particles. Only somewhat later does the number of particles on that
level become substantial. This is why  the evolution of
the force resolution is consistent with nearly constant proper force
resolution of 1~\kpch~ from $z=20$ to $z=0$.

The ART code uses the expansion parameter $a=(1+z)^{-1}$ as the time
variable. Each particle moves with its own time-step which depends on
the refinement level.  The zero-level defines the maximum value $\Delta
a$ of the stepping of the expansion parameter. For Bolshoi $\Delta a
=2\times 10^{-3}$ for $a<0.8$ and $\Delta a =3\times 10^{-3}$ for later
moments. The time-step decreases by a factor of two from one level of
refinement to the next. There were ten levels of refinement at the later
stages of evolution ($z<1$). This gives the effective number of
400,000 time-steps. 

The initial conditions for the simulation were created using the
Zeldovich approximation with particles placed in a homogeneous grid
\citep{Klypin83, Klypin97}. Bolshoi starts at $z_{\rm init}=80$ when
the rms density fluctuation $\Delta\rho/\rho$ in the computational box
with $2048^3$ particles was equal to $\Delta\rho/\rho=0.0826$. (The
Nyquist frequency defines the upper cutoff of the spectrum, with the
low cutoff being the fundamental mode.) The initialization code uses
the Zeldovich approximation, which provides accurate results only if
the density perturbation is less than unity. This should be valid at
every point in the volume. In practice, the fluctuations must be even
smaller than unity for high accuracy. With $2048^3$ independent
realizations of the density, one expects to find one particle in the
box to have a 6.5$\sigma$ fluctuation. For this particle the density
perturbation is $6.5\times 0.0826 = 0.55$; still below 1.0. The most
dense 100 particles are expected to have a density contrast
$5.76\times0.0856 = 0.49$, low enough for the Zeldovich approximation
still to be accurate.

The Bolshoi simulation was run at the NASA Ames Research Center on the 
{\it Pleiades} supercomputer. It used 13,824 cores (1728 MPI tasks each
having 8 OpenMP threads) and a cumulative 13~Tb of RAM. We saved 180 snapshots for
subsequent analysis. The total number of files saved in different
formats is about 600,000, which use 100~Tb of disk space.

For some comparisons we also use a catalog of halos for the Via
Lactea-II simulation \citep{VLII}. This simulation was run for one
isolated halo with maximum circular velocity $\Vcirc =201$~\kms. Using
approximations for the dark matter profile provided by \citet{VLII},
we estimate the virial mass and radius of the halo to be $M_{\rm
  vir}=1.3\times 10^{12}\,\Msunh$ and $R_{\rm vir}=226\,\kpch$. Here we
use the top-hat model with cosmological constant $\Lambda$ to estimate
the virial radius, as explained in the next section.  The Via Lactea-II 
simulation uses slightly different cosmological parameters as
compared with Bolshoi (see Table~1). In particular, the amplitude of
perturbations is 10 percent lower: $\sigma_8=0.74$.

\section{Halo identification}

Let us start with some definitions. 
A distinct halo is a halo that does not ``belong'' to another halo:
its center is not inside of a sphere with a radius equal to the virial
radius of a larger halo. A subhalo is a halo whose center is inside
the virial radius of a larger distinct halo. Note that distinct halos
may overlap: the same particle may belong to (be inside the virial
radius of) more than one halo. We call a halo isolated if there is no
larger halo within twice its virial radius. In some cases we study
very isolated halos with no larger halo within three times its virial
radius.

We use the Bound-Density-Maxima (BDM) algorithm to identify halos in
Bolshoi \citep{Klypin97}. Appendix A gives some of the details of the
halofinder. \citet{Knebe11} present detailed comparison of BDM with
other halofinders and show results of different tests. The code
locates maxima of density in the distribution of particles, removes
unbound particles, and provides several statistics for halos including
virial mass and radius, and maximum circular velocity
\begin{equation}
 \Vcirc=\sqrt{\frac{GM(<r)}{r}}\Big|_{\rm max} .
\end{equation}
Throughout this paper we will use $\Vcirc$ and the term {\it circular
  velocity} to mean maximum circular velocity over all radii $r$. When
a halo evolves over time, its $\Vcirc$ may also evolve. This is
especially important for subhalos, which can be significantly tidally
stripped and can reduce their $\Vcirc$ over time. Thus, we distinguish
instantaneous $\Vcirc$ and the peak circular velocity over halo's
history.

We use the virial mass definition \Mvir~ that follows from the
top-hat model in an expanding Universe with a cosmological
constant. We define the virial radius $R_{\rm vir}$ of  halos as the radius
within which the mean density is the virial overdensity times the mean
universal matter density $\rho_{\rm m}=\Omega_{\rm M}\rho_{\rm crit}$ at that
redshift. Thus, the virial mass is given by
\begin{equation}
    M_{\rm vir} \equiv {{4 \pi} \over 3} \Delta_{\rm vir} \rho_{\rm m} R_{\rm vir}^3 \ .
\end{equation}
 For our set of cosmological parameters, at $z=0$ the virial
radius \Rvir~ is defined as the radius of a sphere with overdensity of
360 of the average matter density. The overdensity limit changes with
redshift and asymptotically goes to 178 for high $z$.  Different
definitions are also found in the literature. For example, the often
used overdensity 200 relative to the {\it critical} density gives mass
$M_{200}$, which for Milky-Way-mass halos is about 1.2-1.3 times
smaller than \Mvir. The exact relation depends on halo concentration.

Overall, there are about 10~million halos in Bolshoi. Halo catalogs
are complete for halos with $\Vcirc>50$~\kms~ ($\Mvir \approx
1.5\times 10^{10}\Msunh$). We track evolution of each halo in time using
 $\sim$180 stored snapshots. The time difference between
consecutive snapshots is rather small: $\sim$(40-80)~Myrs.

\section{Halo mass and  concentration  functions}
\begin{figure*}[tb!]
\epsscale{2.3}
\plottwo{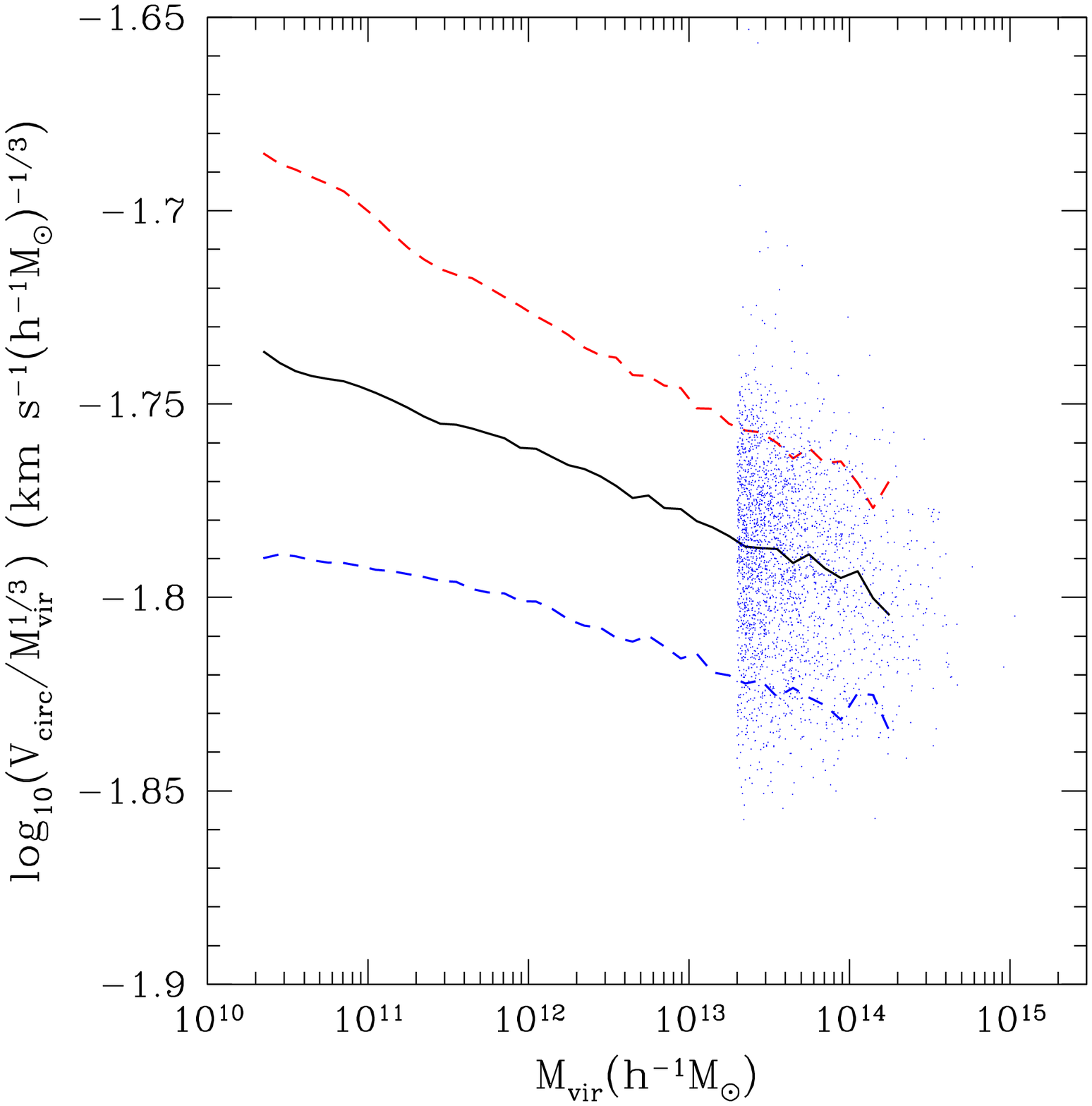}{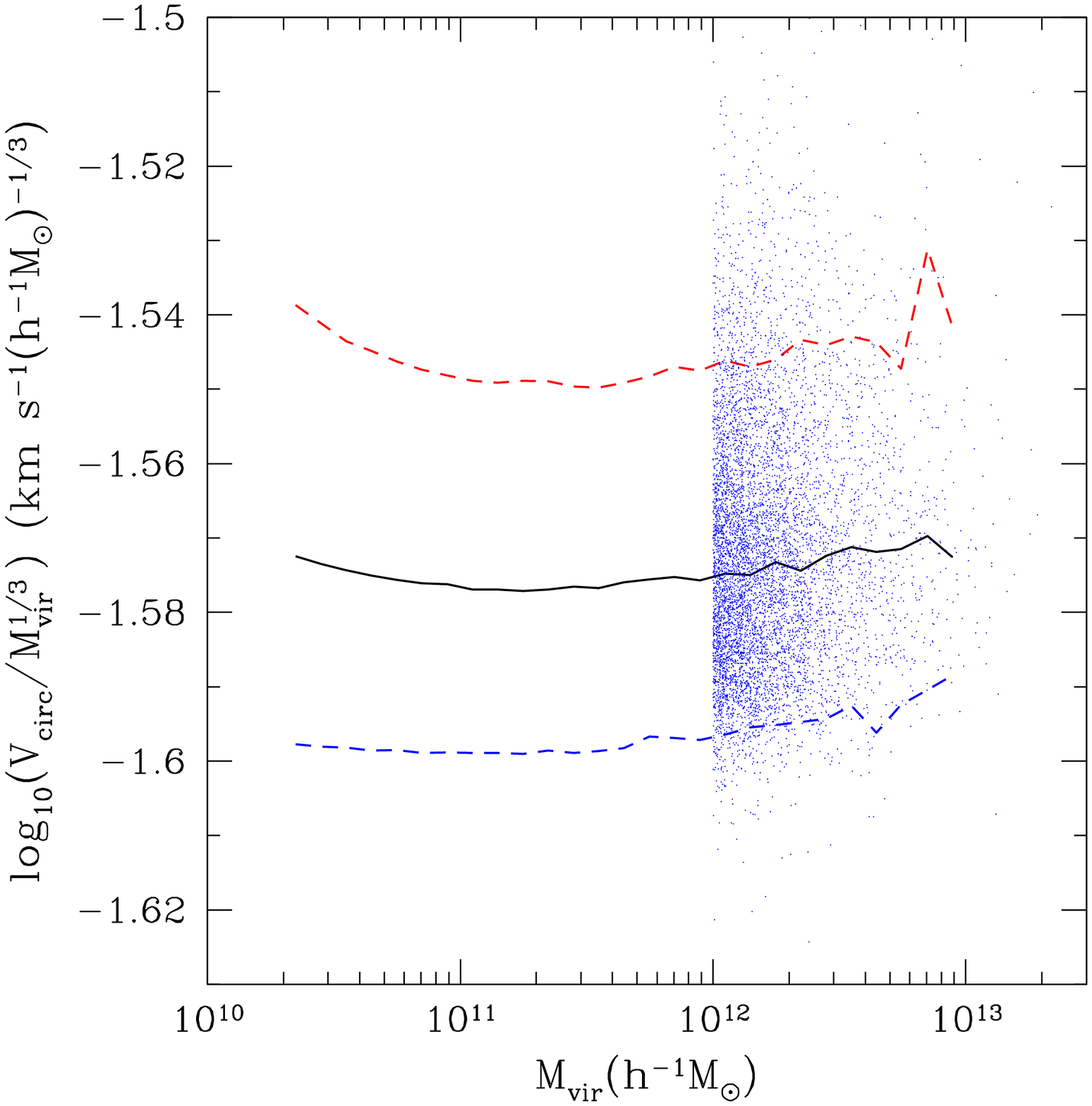}
\caption{Dependence of maximum circular velocity \Vcirc~ on halo mass
  for distinct halos at redshift $z=0$ ({\it left panel}) and redshift
  $z=3$ ({\it right panel}). The circular velocity at any moment
  mostly scales as $\Vcirc\propto \Mvir^{1/3}$. The Figure shows
  deviations from this scaling law. The deviations  are related
  with the halo concentration. Full curves on both plots show median
  \Vcirc~ and dashed curves show 90\% limits.  Dots represent individual halos
  with large masses.}
\label{fig:MVrelation}
\end{figure*}

Throughout most of the paper we characterize halos using their circular
velocity. However, \Vcirc~ tightly correlates with halo mass, as
demonstrated in Figure~\ref{fig:MVrelation}. For distinct halos with
$\Mvir =10^{12}-10^{14}\Msunh$ 90\% of halos have their circular
velocities within 8\% of the median value. Even 99\% of halos are
within only 15-20\%. The variations are substantially larger for
subhalos: 90\% of subhalos with masses $10^{11}-10^{13}\Msunh$ lie
within 20\% of the median \Vcirc.  On average, the circular velocity
increases with mass. The $\Vcirc - \Mvir$ relation depends on halo
concentration $c(\Mvir)$, and, thus, studying this relation gives us a
way to estimate $c(\Mvir)$ without making fits to individual halo
profiles. Any halo mass profile can be parameterized as $M(r)
=M_0f(r/r_0)$, where $M_0$ and $r_0$ are parameters with mass and
radius units and the function $f(x)$ is dimensionless. If
$x_{\rm max}$ is the dimensionless radius corresponding to the maximum
circular velocity, then we can write the following relation between
the maximum circular velocity and the virial mass \citep{Klypin01}:

\begin{eqnarray}
\Vcirc &=& \left[G\frac{f(x_{\rm max})}{f(c)}\frac{c}{x_{\rm max}}
                 \hat\rho^{1/3} \right]^{1/2}\Mvir^{1/3},  \label{eq:vNFW} \\
 \hat\rho &=&\frac{\Mvir}{\Rvir^3} = \frac{4\pi}{3} \Delta_{\rm vir}\rho_{cr}\Omega_{\rm M},\\
f(x) &=& \ln(1+x)-\frac{x}{1+x}, \label{eq:fNFW} \\
c &=& \frac{\Rvir}{r_s}, \quad x =\frac{R}{r_s}, \quad x_{\rm max}=2.15 \label{eq:xNFW}.
\label{eq:VM}
\end{eqnarray}
Here $\Delta_{\rm vir}$ is the overdensity limit that defines the
virial radius; $\rho_{cr}$ and $\Omega_{\rm M}$ are the critical
density and the contribution of matter to the average density of the
Universe. The first two equations are general relations for any
density profile. Eqs.~(\ref{eq:fNFW}) and (\ref{eq:xNFW}) are specific
for NFW: $r_s$ is the characteristic radius of the NFW profile, which
is the radius at which the logarithmic slope of the density is $-2$. At $z=0$, for our
cosmological model, $\Delta_{\rm vir}=360$ and $\Omega_{\rm
  m}=0.27$. Calculating the numerical factors in
eqs.(\ref{eq:vNFW}-\ref{eq:VM}) we get the following relation between
virial mass, circular velocity and concentration at $z=0$:
\begin{equation}
   \Vcirc(\Mvir) = \frac{6.72\times 10^{-3}\Mvir^{1/3}\sqrt{c}}{\sqrt{\ln(1+c)-c/(1+c)}}, 
\end{equation}
where mass $\Mvir$ is in units of $h^{-1}M_\odot$, circular velocity
is in units of \kms. This relation gives us an opportunity to estimate
halo concentration directly for given virial mass and circular velocity.

Alternatively, one may skip the $c(M)$ term and simply use power-law
approximations for the $\Vcirc-\Mvir$ relation which give a good fit
to numerical data.

\noindent For distinct halos:
\begin{equation}
   \Vcirc(\Mvir) = 2.8\times 10^{-2}\Mvir^{0.316}, 
\end{equation}
and for subhalos:
\begin{equation} 
  \Vcirc(\Msub) = 3.8\times 10^{-2}\Msub^{0.305}.
\end{equation}
 Here velocities are in \kms~ and masses are in units
of $\Msunh$. 

Equations (\ref{eq:vNFW} - \ref{eq:VM}) can be considered as
equations for halo concentration: for given $z$, $\Mvir$, and
$\Vcirc$ one can solve them to find $c$. For quiet halos the result
must be the same as what one gets from fitting halo density profiles
with the NFW approximation: two independent parameters (in our case
$\Mvir$ and $\Vcirc$) uniquely define the density profile. However, by
using only two parameters, we are prone to fluctuations. In order to
reduce the effects of fluctuations we apply eqs.~(\ref{eq:vNFW}
- \ref{eq:VM}) to the {\it median} values of $\Vcirc$ for each mass bin.
For each mass bin with the average $\Mvir$~ we find the median
circular velocity $\Vcirc(\Mvir,z)$. We then solve
equations~(\ref{eq:vNFW}-\ref{eq:xNFW}) and get median halo
concentrations $c(\Mvir,z)$. One can also find concentration for each
halo and then take the median - result is the same because for a given
mass the relation between $c$ and $\Vcirc$ is monotonic. This
procedure minimizes effects of fluctuations and gives the median halo
concentration for a given mass. \citet{Munos2010} applied our method to
their simulations and reproduced results of direct density profile
fitting. \cite{Prada11} state that this method recovers results of
halo concentrations $c(M)$ in MS-I \citep{Neto07} with deviations of less than 5
percent over the whole range of masses in the simulation.

\begin{figure}[tb!]
\epsscale{1.0}
\plotone{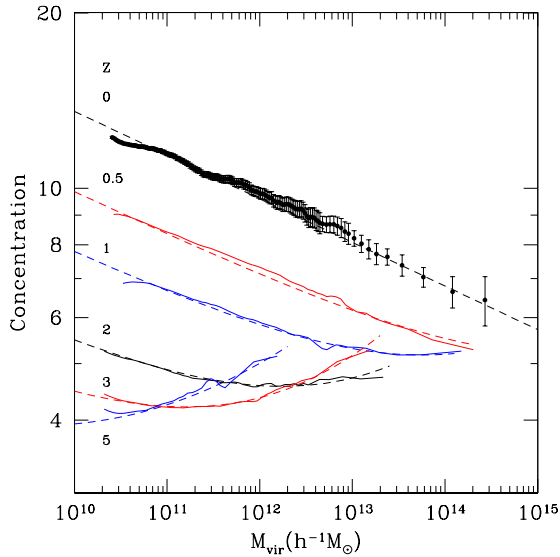}
\caption{Evolution of concentration of distinct halos with
  redshift. The full curves and symbols show results of simulations. Analytical
  approximations are shown as dashed curves. All the fits have the
  same functional form of eq.~(\ref{eq:CM}) with two free parameters. At
  low redshifts the halo concentration decreases with increasing
  mass. However, the trend changes at high redshifts when the
  concentration is nearly flat and even has a tendency to slightly
  increase with mass. }
\label{fig:CMz}
\end{figure}

Figure~\ref{fig:CMz} shows the concentrations for
redshifts $z=0-5$. For redshift zero we get the following approximation:
\begin{equation}
   c(\Mvir) = 9.60\left(\frac{\Mvir}{10^{12}\Msunh}\right)^{-0.075}
\end{equation}
for distinct halos. For subhalos we find:
\begin{equation}
   c(\Msub) = 12\left(\frac{\Msub}{10^{12}\Msunh}\right)^{-0.12}.
\end{equation}

  Subhalos are clearly more concentrated than distinct
halos with the same mass, which is likely caused by tidal
stripping. The differences between subhalos and distinct halos on
average are not large: a 30\% effect in halo concentration.

We also study the evolution of distinct halo concentration with
redshift. Figure~\ref{fig:CMz} shows the concentrations for redshifts
$z=0-5$. Results can be approximated using the following functions:

\begin{eqnarray}
   c(\Mvir,z) &=& c_0(z)\left(\frac{\Mvir}{10^{12}\Msunh}\right)^{-0.075} \nonumber \\
                     &&   \times\left[1+\left(\frac{\Mvir}{M_0(z)}\right)^{0.26} \right],
\label{eq:CM}
\end{eqnarray}
where $c_0(z)$ and $M_0(z)$ are two free factors for each
$z$. Table~\ref{tab:Conc} gives the parameters for this approximation at
different redshifts. For convenience we also give the concentration for
a virial mass $10^{12}\,\Msunh$ and the minimum value of concentration $c_{\rm
  min}$. The simulation box for the Bolshoi is not large enough to find
whether there is a minimum concentration for $z<0.5$. For these
epochs the Table gives the value of concentration at $10^{15}\Msunh$
as predicted by the analytical fits.

\begin{figure}[tb!]
\epsscale{1.0}
\plotone{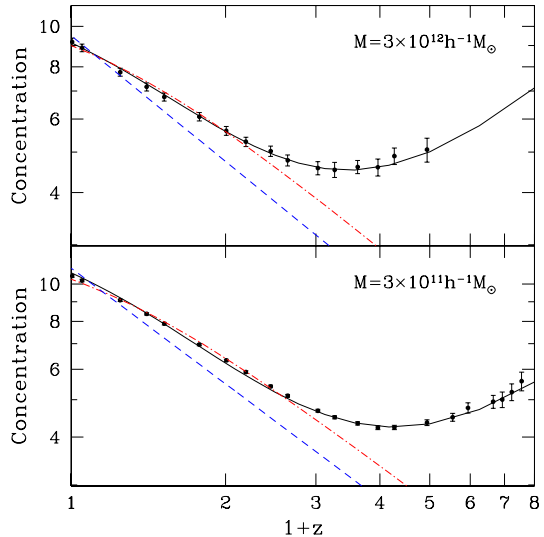}
\caption{Evolution of halo concentration for halos with two masses
  indicated on the plot. The dots show results of
  simulations. For the reference the dashed lines show a power-law
  decline $c\propto (1+z)^{-1}$. Concentrations do not change as fast
  as the law predicts.  At low redshifts $z<2$ the decline in concentration
  is $c\propto \delta$ (dot-dashed curves), where $\delta$ is the
  linear growth factor. At high redshifts the concentration flattens
  and then slightly increases with mass. For both masses the
  concentration reaches a minimum of $c_{\rm min}\approx 4-4.5$, but
  the minimum happens at different redshifts for different masses.
  The full curves are analytical fits with the functional form of
  eq.~(\ref{eq:fitC}). }
\label{fig:Conc}
\end{figure}
The curves in Figure~\ref{fig:CMz} look different for different
redshifts. Typically the concentration declines with redshift and the
shape of the curves evolves.
Another interesting result is that the high-$z$ curves show that the
halo concentration has an upturn: for the most massive halos the
concentration {\it increases} with mass. In order to demonstrate this
more clearly, we study in more detail the evolution with redshift of
the halo concentration for halos with two masses: $3\times
10^{11}\Msunh$ and $3\times 10^{12}\Msunh$. Note that the masses are
the same at different redshifts. So, this is not the evolution of the
same halos. Figure~\ref{fig:Conc} shows the results. Just as expected,
in both cases at low redshifts the halo concentration declines with
redshift. The decline is {\it not} as steep as often assumed $c\propto
(1+z)^{-1}$; it is significantly shallower even at low $z$. For $z<2$
a power-law approximation $c\propto \delta(z)$ is a much better fit,
where $\delta(z)$ is the linear growth factor. It is also a better
approximation because the evolution of concentration should be related
with the growth of perturbations, not with the expansion of the
universe. At larger $z$ the concentration flattens and slightly
increases at $z>3$. The upturn is barely visible for the larger mass,
but it is clearly seen for the $3\times 10^{11}\Msunh$ mass
halos. These and other results show that the concentration in the
upturn does not increase above $c\approx 5$ though it may be related
with the finite box size of our simulation. There is also an
indication that there is an absolute minimum of the concentration
$c_{\rm min}\approx 4$ at high redshifts.  Relaxed halos\footnote{
  Relaxed halos are defined as halos with offset parameter $X_{\rm
    off} < 0.07$ and with spin parameter $\lambda<0.1$, where
  $X_{\rm off}$ is the distance from the halo center to its center of mass
  in units of the virial radius.}  show a slightly {\it stronger
  upturn} indicating that non-equilibrium effects are not the prime
explanation for the increasing of the halo concentration.

 The following analytical approximations provide fits for the evolution of
concentrations for fixed masses as shown in Figure~\ref{fig:Conc}:

\begin{equation}
   c(\Mvir,z) = c(\Mvir,0)
                        \left[\delta^{4/3}(z)+\kappa(\delta^{-1}(z)-1) \right],
\label{eq:fitC}
\end{equation}
here $\delta(z)$ is the linear growth factor of fluctuations normalized to
be $\delta(0) =1$ and $\kappa$ is a free parameter, which for the
masses presented in the Figure is $\kappa=0.084$ for $M=3\times 10^{11}\Msunh$ and $\kappa=0.135$
for ten times more massive halos with $M=3\times 10^{12}\Msunh$.

\begin{table}[htb!]
\begin{center}
\tabcolsep 4.5pt
\caption{Parameters of fit eq.(\ref{eq:CM}) for virial halo
concentration\label{tab:Conc}}
\small
\begin{tabular}{lcccc}
\tableline
\tableline
Redshift & $c_0$ & $M_0/\Msunh$ & $c_{\rm min}$ & $c(10^{12}\Msunh)$\\
\tableline
0.0  & 9.60 & -- & --   & 9.60 \\
0.5  & 7.08 & $1.5\times 10^{17}$ & 5.2   & 7.2 \\
1.0  & 5.45 & $2.5\times 10^{15}$ & 5.1   & 5.8 \\
2.0  & 3.67 & $6.8\times 10^{13}$ & 4.6   & 4.6 \\
3.0  & 2.83 & $6.3\times 10^{12}$ & 4.2   & 4.4 \\
5.0  & 2.34 & $6.6\times 10^{11}$ & 4.0   & 5.0 \\
\tableline
\end{tabular}
\end{center}
\end{table}

\begin{figure}[tb!]
\epsscale{1.0}
\plotone{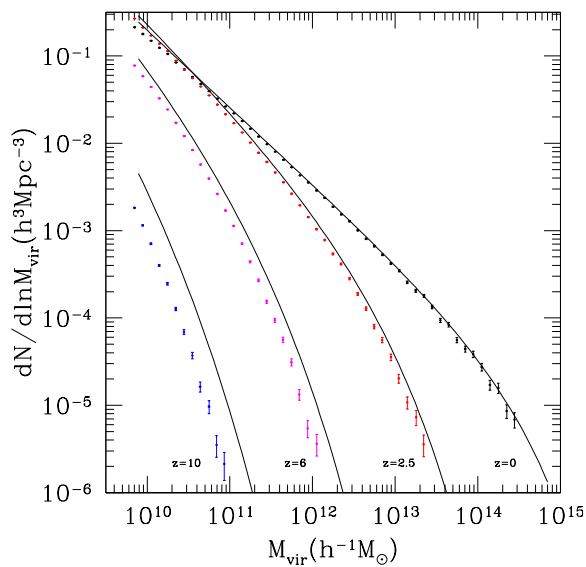}
\caption{Mass function of distinct halos at different redshifts
  (circles). Curves show the Sheth-Tormen approximation, which
  provides a very accurate fit at $z=0$, but overpredicts the number
  of halos at higher redshifts.}
\label{fig:MassFunction}
\end{figure}

It is interesting to compare these results with other simulations.
\citet{Zhao03,Zhao09} were the first to find that the concentration
flattens at large masses and at high redshifts. Their estimates of the
minimum concentration are compatible with our results. Figure 2 in
\citet{Zhao03} shows an upturn in concentration at $z=4$. However, the
results were noisy and inconclusive: the text does not even mention
it.

\citet{Maccio08} present results that can be directly compared with
ours because they use the same definition of the virial radius and
estimate masses within spherical regions. Their models named WMAP5
have parameters that are very close to those of Bolshoi. There is one
potential issue with their simulations.  \citet{Maccio08} use a set of
simulations with each simulation having a small number of particles
and either a low resolution, if the box size is large, or very small
box, if the resolution is small. For all the halos in their simulations
the approximation for concentration is $c(\Mvir) =
8.41(\Mvir/10^{12}\Msunh)^{-0.108}$. Bolshoi definitely gives more
concentrated halos. The largest difference is for cluster-size
halos. For $\Mvir =10^{15}\Msunh$ our results give $c=5.7$ while
\citet{Maccio08} predict $c=4.0$ -- a 40\% difference. The difference
gets smaller for galaxy-size halos: 14\% for $\Mvir =10^{12}\Msunh$
and 2\% for $\Mvir =10^{10}\Msunh$. Comparing results for
relaxed halos we find that the disagreement is smaller. \citet{Munos2010}
give $c(\Mvir) =9.8$ for $\Mvir =10^{12}\Msunh$ as compared with our
results (also for relaxed halos) of $c(\Mvir) =10.1$ - a 3\%
difference.  For clusters with $\Mvir =10^{15}\Msunh$ the disagreement
is 18\%.

\begin{figure*}[tb!]
\epsscale{2.35}
\plottwo{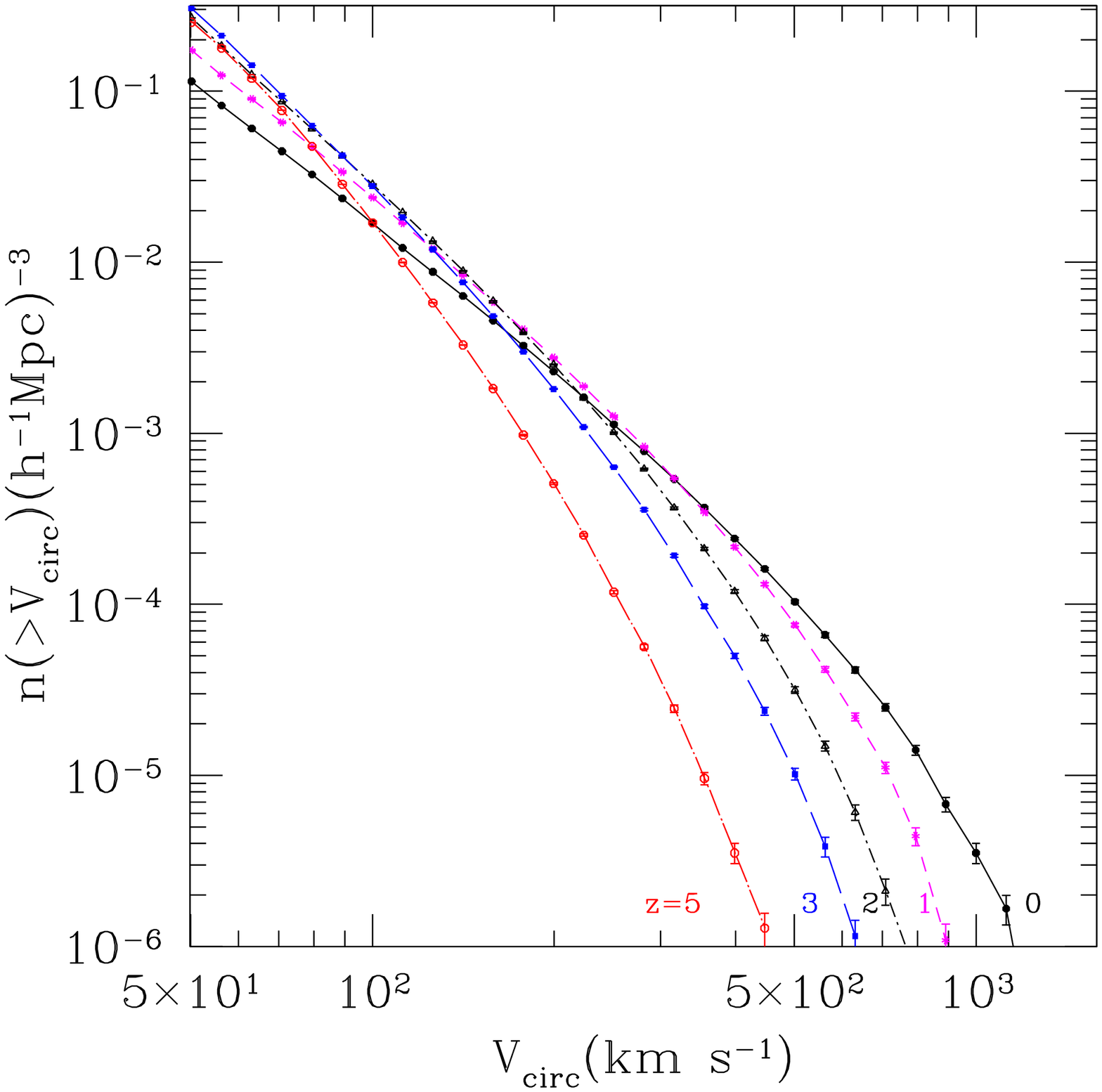}{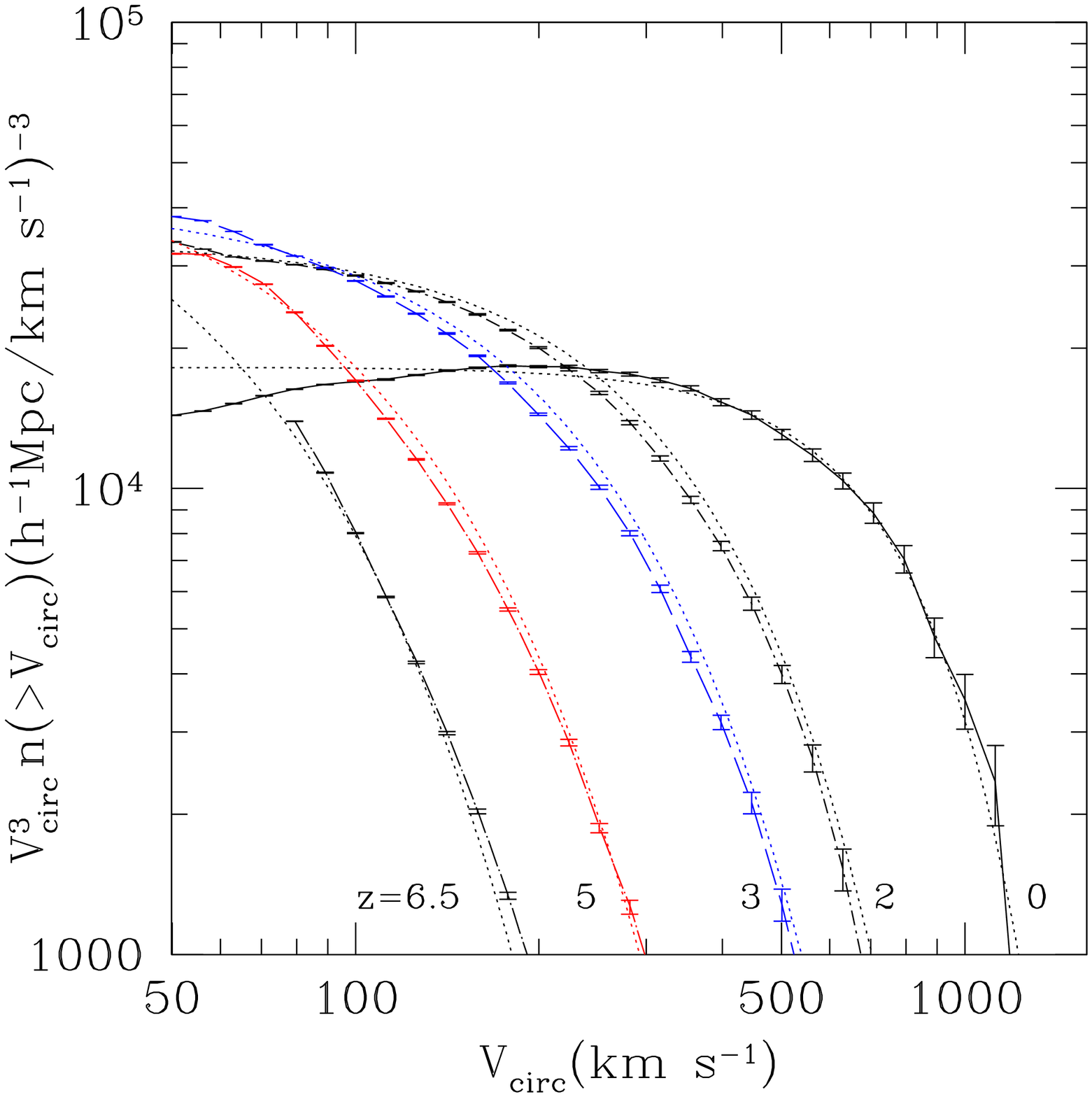}
\caption{Velocity function of distinct halos at different
  redshifts. {\it Left:} Symbols and curves show the cumulative velocity
  function of the Bolshoi simulation. Error bars show Gaussian
  fluctuations. {\it Right:} We plot the product $V^3n(>V)$ of the
  cumulative velocity function and the circular velocity.  Dotted
  curves show analytical approximations with the form given by 
  $V^3n(>V)=A\exp(-[V/V_0]^\alpha$), which provide accurate fits to
  numerical results.}
\label{fig:VelFunction}
\end{figure*}

We can also compare our results with those of MS-I \citep{Neto07}
though MS-I has different cosmological parameters and a different power
spectrum.  Because the  analysis of MS-I was done for the overdensity
200, we also made halo catalogs for this definition of halos.
\citet{Neto07} give the following approximation for all halos:
$c_{200} = 7.75(M_{200}/10^{12}\Msunh)^{-0.11}$. 
For halos in the Bolshoi simulation $c_{200} = 7.2(M_{200}/10^{12}\Msunh)^{-0.075}$.
 Thus the MS-I $c_{200}$ is 8\% larger than the 
concentrations in Bolshoi for $\Mvir=10^{12}\Msunh$, with a small ($\sim $10\%)
difference for $\Mvir=$ $10^{14}-10^{15}\Msunh$.  For $\Mvir=$ $10^{12}\Msunh$ in
the MS-II and Aquarius simulations \citet{Boylan09} give an even larger
concentration of $c_{\rm vir}=12.9$, which is 1.3 times larger
than what we get from Bolshoi.  Most of the differences are likely due
to the larger amplitude of cosmological fluctuations in MS simulations
because of the combination of a larger $\sigma_8$ and a steeper spectrum
of fluctuations.

The mass function of distinct halos is a classical cosmological result
\citep[e.g.,][]{Warren06,Reed07,Tinker08,Maccio08,Reed09}.
Figure~\ref{fig:MassFunction} presents the results for the Bolshoi
simulation together with the predictions of the Sheth-Tormen
approximation \citep[][ST, see also Appendix B]{ST02}. We find that the
ST approximation gives an accurate fit for $z=0$ with the
deviations less than 10 percent for masses ranging from $\Mvir=5\times
10^{9}\Msunh$ to $\Mvir=5\times 10^{14}\Msunh$. However, the ST
approximation overpredicts the abundance of halos at higher redshifts.
For example, at $z=6$ for halos with $\Mvir\approx (1-10)\times
10^{11}\Msunh$ the ST approximation gives a factor of 1.5 more halos
as compared with the simulation. At redshift ten the ST approximation
gives a factor of ten more halos than what we find in the simulation.

We introduce a simple correction factor which brings the analytical
predictions much closer to the results of simulations. We find that
the ST approximation multiplied by the following factor gives 
less than 10 percent deviations for masses $5\times 10^{9}\Msunh - 5\times
10^{14}\Msunh$ and redshifts $z=0-10$:

\begin{equation}
   F(\delta) = \frac{(5.501\delta)^4}{1+(5.500\delta)^4},
\label{eq:STcorrect}
\end{equation}
where $\delta$ is the linear growth rate factor normalized to unity at
$z=0$ (see eq.~(\ref{eq:delta})).

Our results are in good agreement with \citet{Tinker08}, who present
the evolution of the mass function for $z=0-2.5$ for halos defined
using the spherical overdensity method.  At redshift zero, their mass
function for overdensity $\Delta =200$ relative to the mean mass
density is 20\% above the ST approximation. This is expected because
masses defined with spherical overdensity $\Delta =200$ are typically
10-20\% larger than virial masses, which are used in our paper. Results in
\citet{Tinker08} together with our work indicate that at higher
redshifts the mass function gets more and more below the ST
approximation. In addition, the shape of the mass function in
simulations gets steeper: there is a larger disagreement at larger
masses.  \citet{Tinker08} argue that this behavior indicates that the
mass function is not ``universal'': it does not scale with the
redshift only as a function of the amplitude of perturbations
$\sigma(M)$ on scale $M$ (see Appendix B for definitions).  Our results
extend this trend to redshifts at least as large as $z=10$. Our
results also qualitatively agree with \citet{Cohn08}, who present spherical
overdensity halo masses at $z=10$. They also find substantially lower
mass functions as compared with the ST approximation, though the
differences with the approximation are somewhat smaller than what we
find.

\begin{figure}[tb!]
\epsscale{1.0}
\plotone{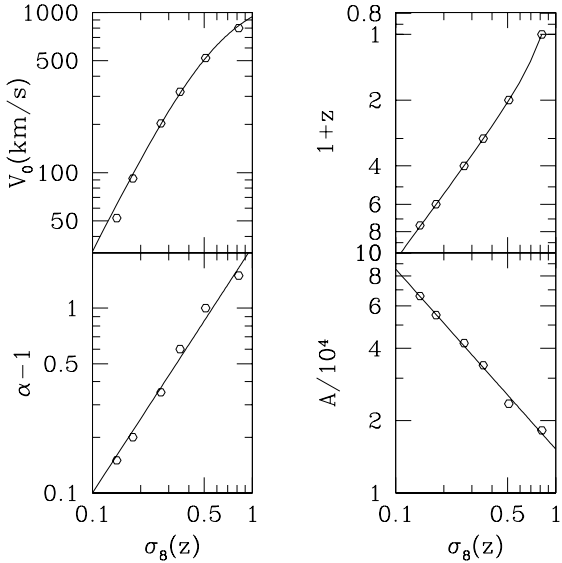}
\caption{Parameters of the velocity function at different amplitudes of
  perturbations $\sigma_8(z)$. Open circles show parameters found by
  fitting $n(>V,z)$ at different redshifts. The curves
  are power-law fits given by eqs.~(\ref{eq:nv}-\ref{eq:pars}). The
  top right panel shows the evolution of $\sigma_8$ with redshift as
  predicted by the linear theory. Circles on the curve indicate
  the same moments as on the other three panels.}
\label{fig:VelFpars}
\end{figure}

These results are at odds with those obtained with the
Friends-Of-Friends (FOF) method
\citep[e.g.][]{Lukic07,Reed07,Reed09,Cohn08}. The FOF halo mass
function scales very close to the ``universal'' $\sigma(M)$ behavior.
The reason for the disagreement between spherical overdensity and FOF
halo finding methods is likely related with the fact that FOF has a
tendency to link together structures before they become a part of a
virialized halo. This happens more often with the rare most massive
halos, which have a tendency to be out of equilibrium and in the
process of merging.  As a result of this, FOF masses are artificially
inflated.  \citet{Cohn08} studied case-by-case some halos at $z=10$
and conclude that in their simulations FOF assigned to halos almost twice more
mass. Comparison of the ST predictions with the Bolshoi results shown
in Figure~\ref{fig:MassFunction} points to the difference of a factor
of 2.5 in mass for $z=10$. Because of the steep decline of the mass function,
a factor of 2.5 increase in mass translates to a factor of ten
increase in the number-density of halos. This correction to the FOF
masses must be taken into account when making any estimates of the
frequency of appearance of high-$z$ objects.

In Appendix C we also directly compare FOF masses with those obtained
with the BDM code. At $z=8.8$ the FOF masses with the linking
parameter $l=0.20$ were on average 1.4 times larger than the BDM
masses. In addition, the spread of estimates was very large with FOF in
many cases giving 3-5 times larger masses than BDM. Analysis of
individual cases shows that this happens because FOF links large
fragments of filaments, not just an occasional neighboring halo.  The
situation is different at $z=0$. Here both BDM and FOF ($l=0.17$) give
remarkably similar results, though some spread is still present
\citep{Tinker08}. We speculate that the difference in the behavior at
high and low $z$ is related with the slope of the power spectrum of
perturbations probed by halos at different redshifts.

These differences between different definitions of masses and radii of
halos indicate the inherent weakness of masses as halo properties: in
the absence of a well defined physical process responsible for halo
formation, masses are defined somewhat arbitrarily. We know that halos
do not form according to the often used top-hat model. We also
know that the virial radius is ill-defined for non-isolated
interacting objects. Nevertheless, we use one definition or
another and we pay a price for this vagueness. These uncertainties in
masses also motivate us to use another, -- much better defined quantity
-- the maximum circular velocity.

\section{Halo velocity function}

The velocity function for distinct halos is shown in
Figure~\ref{fig:VelFunction}. It declines very steeply with velocity.
At small velocities the  power slope is --3  with an exponential
cutoff at large velocities. We find that at all redshifts the
cumulative velocity function can be accurately approximated by the
following expression:

\begin{equation}
   n(>V)  = AV^{-3}\exp\left(-\left[\frac{V}{V_0} \right]^\alpha\right),
\label{eq:nv}
\end{equation}
where the  parameters $A$, $V_0$, and $\alpha$ are functions of
redshift. For $z=0$ we find
\begin{eqnarray}
   A&=&1.82\times 10^4 (\Mpch/\kms)^{-3}, \nonumber \\
   \alpha &=& 2.5, \\
   V_0 &=& 800~\kms. \nonumber
\end{eqnarray}
 The evolution of the
parameters should not directly depend on the redshift, but on the
amplitude of perturbations. Indeed, when we plot the parameters as
functions of $\sigma_8(z)$ as predicted by the linear theory at
different redshifts, the functions are very close to power-laws as
demonstrated by Figure~\ref{fig:VelFpars}.  We find the following fits
to the parameters:
\begin{eqnarray}
A &=& 1.52\times 10^4\sigma_8^{-3/4}(z)(h^{-1}{\rm Mpc}/\kms)^{-3}, \nonumber\\
\alpha &=& 1+ 2.15\sigma_8^{4/3}(z), \label{eq:pars} \\
V_0 &=& 3300 \frac{\sigma_8^{2}(z)}{1+2.5\sigma_8^{2}(z)} \kms. \nonumber
\end{eqnarray}

\section{Abundance of Subhalos}
\begin{figure}[tb!]
\epsscale{1.0}
\plotone{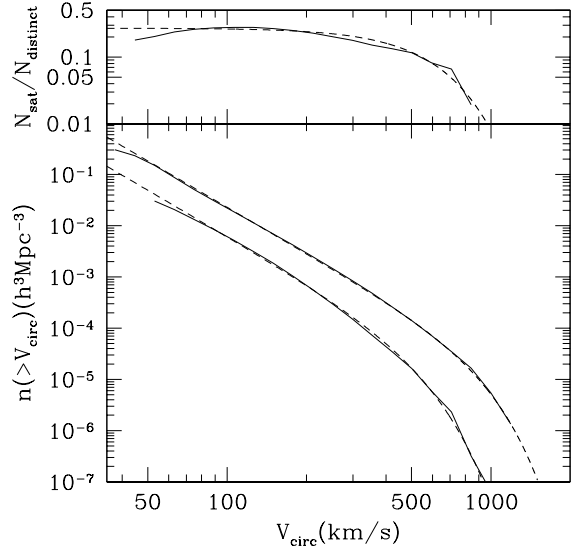}
\caption{The velocity function of satellites compared with the
  velocity function of distinct halos. The bottom panel shows the
  cumulative function of subhalos (bottom full curve) and distinct
  halos (top full curve). The circular velocity used for the plot is
  the peak over each halo's history. The dashed curves are analytical
  approximations. The top panel shows the ratio of the number of
  subhalos and distinct halos (full curve) and an analytical
  approximation for the ratio (dashed curve). }
\label{fig:VsatAll}
\end{figure}
Figure~\ref{fig:VsatAll} shows the cumulative velocity function
$n(>\Vcirc)$ of all subhalos in Bolshoi regardless of the circular
velocity of their host halos. We use maximum circular velocities over
the whole evolution for both subhalos and distinct halos. The top
panel shows the ratio of the number of subhalos to the number of
distinct halos with the same limit on the circular velocity. Note that
for given $\Vcirc$ most of the halos are distinct. This may sound a
bit counter intuitive. Because each distinct halo has many subhalos,
one would naively expect that there are many more satellites as
compared with distinct halos. This is not true. The number of satellites is
large, but  most are small. When we count small distinct halos,
their number increases very fast and we always end up with more
distinct halos at a given circular velocity. The abundance of
subhalos can be approximated with the same function given by eq.~(\ref{eq:nv})
as for the distinct halos. However, the parameters of the approximation
are different. For subhalos that exist at $z=0$ and for which we use
peak velocities over their history of evolution, we find:

\begin{eqnarray}
   A &=& 6.2\times 10^3 (\Msunh/\kms)^{-3} \nonumber \\ 
   \alpha &=&2.2, \quad V_0 = 480\, {\rm \kms}.
  \label{eq:psub}
\end{eqnarray}
The remarkable similarity of the shapes of the velocity functions of halos
and subhalos suggests a simple interpretation for the difference in
their parameters: subhalos were typically accreted at the epoch when
the velocity function of distinct halos had the same parameters
$\alpha$ and $V_0$ as the velocity function of subhalos at
present. For the parameters given by eqs.~(\ref{eq:pars},\ref{eq:psub}) we
get a typical accretion redshift $z_{\rm acc}\approx 1$.

\begin{figure}[tb!]
\epsscale{1.0}
\plotone{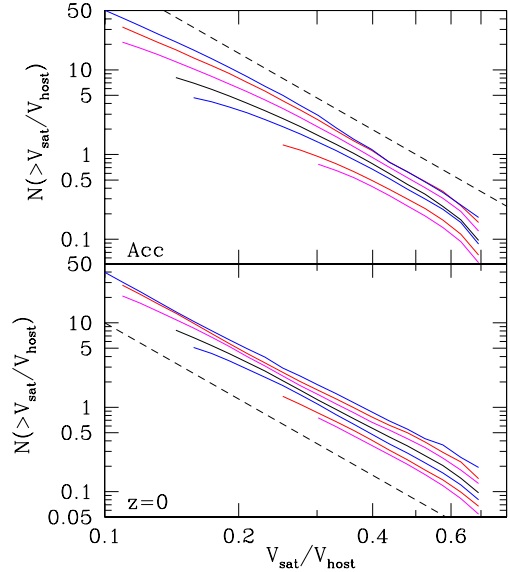}
\caption{The cumulative velocity function of satellites for host halos
  with different maximum circular velocities ranging from $\approx
  150$~\kms~ to $\approx 1000$~\kms~ from bottom to top. The bottom
  panel uses the velocities of subhalos at redshift $z=0$. The top
  panel uses peak circular velocities over the history of each
  subhalo. The dashed lines show power-laws with the slope
  $-3$. The abundance of subhalos increases with increasing host halo mass.}
\label{fig:Vsat}
\end{figure}

In order to study statistics of subhalos belonging to different parent
halos we split our sample of distinct halos into subsamples with
different ranges of circular velocities.  For each subhalo we use
either its $z=0$ circular velocity or the peak value over its entire
evolution.  Figure~\ref{fig:Vsat} shows both the present day and the
peak velocity distribution functions for distinct halos with masses
and velocities ranging from galaxy-size halos to clusters of
galaxies. The average circular velocities for each bin presented in
the figure are (from bottom to top): $V_{\rm host} = (163, 190, 235,
340, 470, 677, 936)$ \kms.  The number of halos in each bin varies from
200 for the most massive halos ($V_{\rm host}=800-1200\, \kms$) to
30000 for the least massive halos with $V_{\rm host}=160-180\,
\kms$. The increase in the abundance of substructure for more massive
hosts is consistent with the results of \citet{Gao04}, who give a factor
of 2.0-2.5 increase for host halos from mass $\sim 2.5\times
10^{12}\Msunh$ to $\sim 10^{15}\Msunh$. \citet{Frank05,Taylor05,Zentner05} came to
similar conclusions using their (semi)analytic models.

\begin{figure}[tb!]
\epsscale{1.0}
\plotone{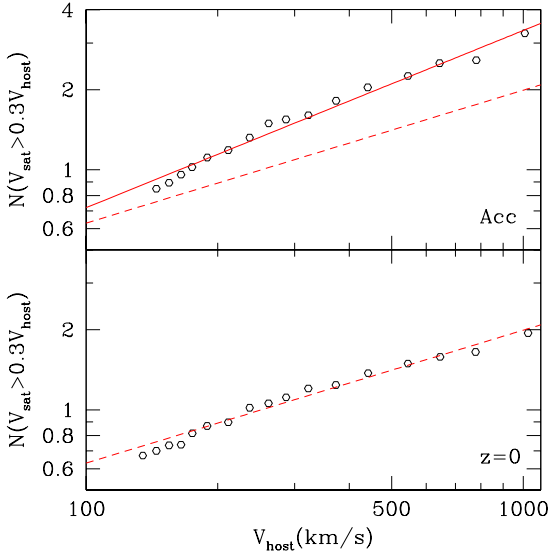}
\caption{Dependence of the number of subhalos on the circular velocity
  of their hosts. Here we count all subhalos with circular velocities
  larger than 0.3 of their host circular velocity. The bottom panel
  shows results for \Vcirc~ estimated at $z=0$ and the top panel is
  for the peak \Vcirc~ over the history of each subhalo. For $z=0$
  circular velocities the abundance scales as $V_{\rm host}^{1/2}$
  (dashed curve). For peak circular velocities the number of subhalos
  is larger and the scaling is steeper: $N\propto V_{\rm host}^{2/3}$
  (full curve). For comparison, the dashed curve is the same as on the
  bottom panel.}
\label{fig:Nscale}
\end{figure}

In order to more accurately measure the dependence of the abundance of
subhalos on the circular velocity of the host halo, we analyze the
number of satellites with circular velocities larger than 0.3 of the
circular velocity of their hosts: $V_{\rm sat}> 0.3V_{\rm host}$. This
approximately corresponds to the mass ratio of $M_{\rm sat}/M_{\rm
  host}\approx 0.3^3 \approx 0.027$.  The threshold of 0.3 is a compromise between
the statistics of satellites and the numerical resolution of the
simulation. Figure~\ref{fig:Nscale} shows the number of satellites
$N_{0.3}(V_{\rm host})$ for hosts ranging from $V_{\rm
  host}\sim$150~\kms~ to $\sim$1000~\kms. The number of satellites
scales as a power-law $N_{0.3}\propto V_{\rm host}^\gamma$ with the
slope $\gamma$ depending on how the circular velocity is estimated. For the
$z=0$ velocities the slope is $\gamma =1/2$, and it is larger for the
peak velocities: $\gamma=2/3$.

\begin{figure}[tb!]
\epsscale{1.0}
\plotone{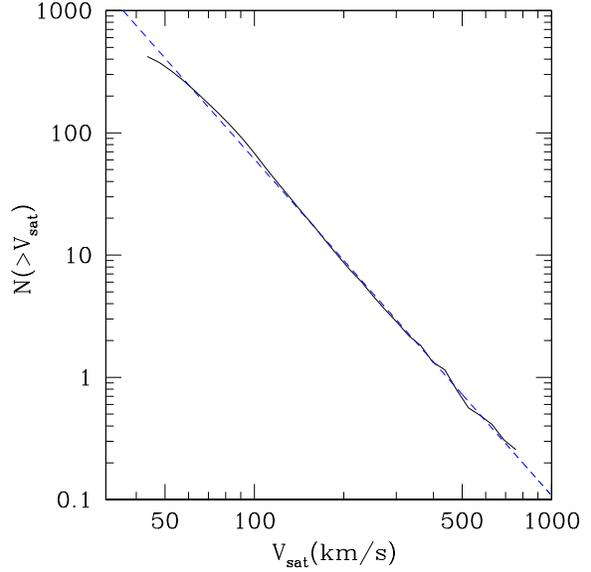}
\caption{Velocity function of subhalos for 40 most massive clusters
  with average $\Mvir=3.2\times 10^{14}\Msunh$ and circular velocity
  $\Vcirc=1100$~\kms. The velocity function is nearly a power law with
  the slope -2.75.}
\label{fig:Vcluster}
\end{figure}

There is an indication that the dependence of the cumulative number of
satellites on their circular velocity $N(>V_{\rm sat})$ gets slightly
shallower for more massive host halos. Figure~\ref{fig:Vcluster}
illustrates the point. Here we study the most massive (but also rare)
halos. Again, the number of satellites is approximated by a
power-law. However, the slope is about $-2.75$, which is somewhat shallower
than the slope $-3$ found for smaller host halos.

\begin{figure}[tb!]
\epsscale{1.0}
\plotone{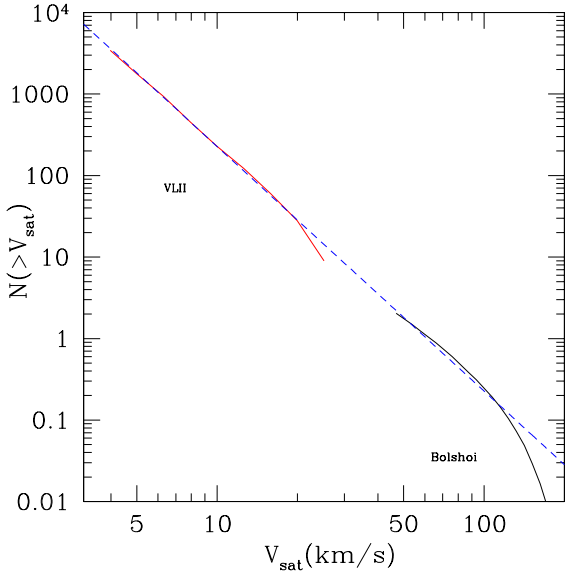}
\caption{Comparison of satellite velocity functions in the Via Lactea-II
  and Bolshoi simulations for host halos with $\Vcirc =200$~kms/s and
  $\Mvir\approx 1.3\times 10^{12}\Msunh$. The dashed line is a power
  law with slope $-3$ which provides an excellent fit to both
  simulations. In both simulations satellites are found inside a sphere
  with virial radius $R_{\rm vir}$.}
\label{fig:VL2}
\end{figure}

We compare some of our results with the Via Lactea-II (VL-II) simulation
\citep{VLII}. We do not use the published results because the 
analysis of VL-II was done using overdensity 180 relative to the matter
density, which gives a  larger radius for halos as compared with the
virial radius. We use the halo catalog of VL-II, which lists coordinates
and circular velocities of individual halos. We also use published
parameterization of the dark matter density in order to estimate the
virial radius of VL-II. When comparing with VL-II, we select halos in
Bolshoi in a narrow range of circular velocities $\Vcirc
=195-205$~\kms. There are 4960 of those with the average virial mass
of $\Mvir =1.26\times 10^{12}\Msunh$, which is close to the virial
mass $M_{\rm vir}=1.3\times 10^{12}\Msunh$ of Via Lactea
II. Figure~\ref{fig:VL2} presents results of the velocity function of
subhalos in those host halos. The dashed line in the figure is the power law
$N(>V)=(V/61{\rm \kms})^{-3}$, which gives a good fit to the data for
a wide range of circular velocities from 4~\kms~ to 100~\kms. 
Bolshoi has slightly more subhalos by about 10\%. This
is a small difference and it goes in line with expectation that
a smaller normalization of cosmological fluctuations  gives fewer
subhalos. In the same vein, the Aquarius simulations have an even higher
(by 30\% as compared to VL-II) number of subhalos, probably because of
an even larger amplitude.

Summarizing all the results, we conclude that the
cumulative velocity function of $z=0$ subhalos can be reasonably
accurately approximated by the power law:
\begin{eqnarray}
N(>x) &=& 1.7\times 10^{-3}V_{\rm host}^{1/2}x^{-3},\\
x &\equiv& V_{\rm sub}/V_{\rm host},\, x<0.7,
\label{eq:Vsub}
\end{eqnarray}
\noindent where the circular velocity of the host is given in units of
\kms.  Again, these results are broadly consistent with the $N$-body
simulations of \citet{Gao04} and with the semianalytic models of
\citet{Frank05,Taylor05} and \cite{Zentner05}.  For peak circular
velocities we obtain:
\begin{equation}
N(>x) = 9.0\times 10^{-4}V_{\rm host}^{2/3}x^{-3}.
\label{eq:Vsubacc}
\end{equation}

\section{The spatial distribution of satellites}

\begin{figure}[tb!]
\epsscale{1.0}
\plotone{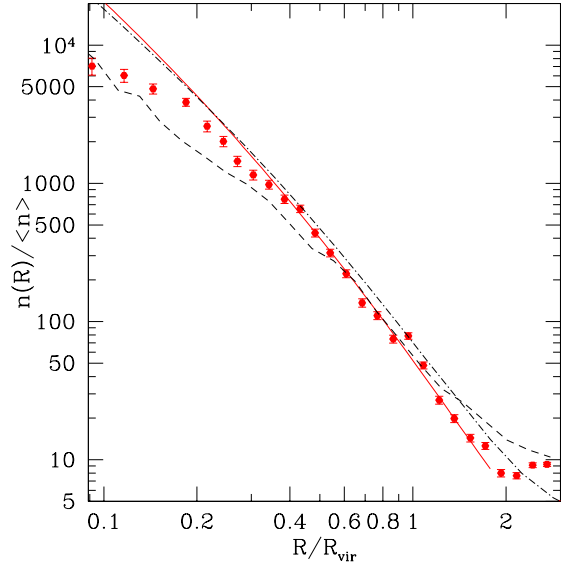}
\caption{Density profiles for galaxy-size halos with $\Vcirc\approx
  200$~\kms. The full curve and circles are the dark matter density
  and the number-density of satellites with $\Vcirc>4$~\kms~ in the
  Via Lactea-II simulation normalized to the average (number-)density
  for each component respectively. The satellites have nearly the same
  overdensity as the dark matter for radii $R=(0.3-2)R_{\rm vir}$. The
  number-density of satellites falls below the dark matter at smaller
  radii.  The dashed curve is the number-density of satellites with
  $\Vcirc>80$~\kms~ found at $z=0$ in the Bolshoi simulation for host
  halos selected to have the same circular velocity as Via
  Lactea-II. In the outer regions with $R=(0.5-1.5)R_{\rm vir}$ the
  satellites follow the dark matter very closely. In the inner regions
  the Bolshoi results are 20-30\% below the much higher resolution
  simulation Via Lactea-II, presumably because of numerical
  effects.}
\label{fig:VLprofile}
\end{figure}

The spatial distribution of satellites has numerous astrophysical
applications. Among others, these include the survivability of dark matter
subhalos \citep[e.g.,][]{Moore96,Klypin99,Colin99}, the potential
annihilation signal of dark matter
\citep[e.g.,][]{Kuhlen08,Springel08,Ando09}, and motion of satellites
as a probe for masses of isolated galaxies and groups
\citep[e.g.,][]{Prada03,Klypin09Sat,More09}. The relative abundance of
satellites and dark matter is a form of bias. Thus, studying the
distribution of satellites in simulations sheds light on the physics
of bias and, thus, on the formation of dwarf galaxies. There is an
additional reason to study the satellites in the  Bolshoi simulation:
comparison with high resolution simulations such as Via Lactea
gives an additional test on resolution effects and provides limits of
the applicability of the simulation.

The spatial distribution of satellites has been extensively studied in
simulations
\citep{Ghigna98,Ghigna00,Nagai05,VLII,Aquarius,Angulo09}. One of the
main issues regarding satellites is to what degree their distribution
is more extended than that of the dark matter. As a small halo falls
into the gravitational potential of a larger halo, it experiences
tidal stripping and dynamical friction.  It may also experience
interaction with other subhalos before and during infall. Tidal
stripping reduces the mass of subhalos, resulting in a very strong
radial bias: subhalos selected by mass have relatively low
number-density in the central region of their hosts. However,
stripping affects much less the central parts of subhalos. This is why
the distribution of subhalos is more concentrated when selected by
their circular velocity \citep{Nagai05}. Depending on the mass and
concentration of the subhalo and on its trajectory, the role of
different physical effects may vary. Interplay of these processes
results in a complicated picture of the distribution of the
satellites. Numerical effects add to the complexity of the situation:
it is a challenge to preserve and to identify small subhalos through
the whole history of evolution of the Universe.

The traditional way of displaying results is to normalize both the
dark matter and satellites to the average mass inside the virial
radius.  When presented in this way, results routinely show that there
are relatively more satellites in the outer parts of halos. For
example, \citet{Angulo09} find that independently of subhalo mass,
subhalos are a factor of two more abundant than dark matter around the
virial radius of their hosts. If that were correct, this would imply
some kind of physical mechanism to produce more satellites outside of
the virial radius, a far-reaching conclusion. However, below we show
that this not correct.

\begin{figure}[tb!]
\epsscale{1.0}
\plotone{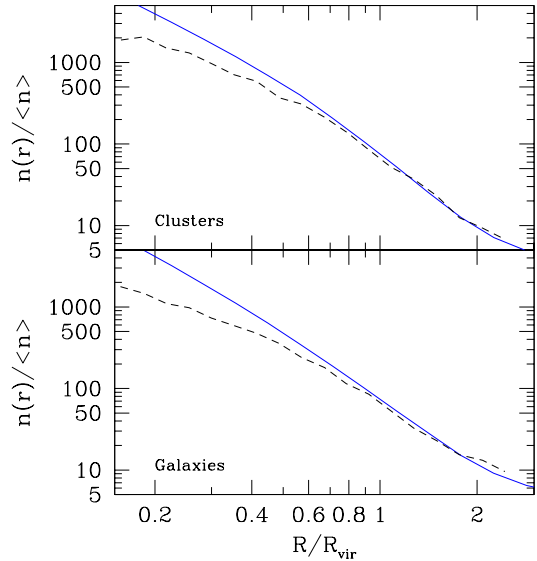}
\caption{Comparison of satellites (dashed curves) and dark matter
  density (full curves) radial distributions for halos with
  $\Mvir=5\times 10^{12}\Msunh$ (bottom panel) and $\Mvir=3\times
  10^{14}\Msunh$ (top panel). Halos were selected to be isolated: no
  other larger halo within radius $2R_{\rm vir}$. Subhalos fallow
  very closely the dark matter density in the outer regions
  $R>0.5R_{\rm vir}$. In the central regions of the halos the
  overdensity of satellites is smaller than that of dark matter. }
\label{fig:SatProfiles}
\end{figure}

The main issue here is the normalization. If a simulation includes
only a small region, a typical setup for modern high-resolution
simulations, there is no sensible way to normalize satellites.  This
is not the case given the statistics of Bolshoi. We can reliably
normalize the abundance of even halos  with small mass and circular
velocities.  The velocity function of all distinct halos is given by
eq.~(\ref{eq:nv}).  There are 20-25\%  subhalos at given 
circular velocity.  Using the velocity function and the fraction of
subhalos from Bolshoi, we estimate the average number of all halos
with any given $\Vcirc$.  These estimates are used to normalize the
number-density profile of subhalos in VL II presented in
Figure~\ref{fig:VLprofile}. A comparison with the dark matter profile
is quite interesting: there is very little bias in the Via Lactea
distribution of subhalos for radii $R=(0.3-2)R_{\rm vir}$. Subhalos
are {\rm not} more extended as compared with the dark matter. However,
Via Lactea-II is just one halo and there may be some effects related with
cosmic variance. We use halos in Bolshoi that  have similar
characteristics as Via Lactea-II: very isolated halos (no equal mass halo inside
$3R_{\rm vir}$) and circular velocities in the range
$\Vcirc=(200-220)$~\kms~ with corresponding masses
$\Mvir=(1.3-1.5)\times 10^{12}\Msunh$. For Bolshoi halos we find a
small $\sim$10\% antibias. In the inner regions ($R<0.5R_{\rm vir}$)
the number density of satellites goes below the results of Via Lactea,
but the difference is not large: 20-30\%.  Some of the differences
with Via Lactea-II may be real because subhalos in Bolshoi are more
massive, and, thus, they must experience stronger
dynamical friction. However, it is more likely that most of the
differences are numerical: after all, Bolshoi has substantially worse
resolution than Via Lactea-II. Regardless of the cause of those small
differences, it is quite remarkable that simulations with five orders
of magnitude difference in mass resolution produce results that
deviate only by 10-20\%.

Comparison with Via Lactea-II is difficult because for these masses
($\Mvir \approx 10^{12}\Msunh$) Bolshoi has only a few subhalos per each
host. In order to have a better picture of the spatial distribution of
satellites, we study more massive halos for which our resolution is
relatively better. Again, we select isolated halos: those with no larger halo
within twice the virial radius. Figure~\ref{fig:SatProfiles} shows
the results for hosts with very different masses. The top panel shows
results for 82 halos with circular velocities in the range
900-1100~\kms~ (average virial mass $2.5\times 10^{14}\Msunh$).  The
bottom panel shows 2200 halos with $\Vcirc=280-300)$~\kms~ and average
$\Mvir =5\times 10^{12}\Msunh$.  Results are very similar for such
different host halos: satellites follow the dark matter very closely for
radii $R=(0.5-2)R_{\rm vir}$ with possible small ($\sim$10\%) antibias. In
the central region  the subhalo abundance goes below the
dark matter by a factor of 2-2.5 at $R=0.2R_{\rm vir}$. 

It is interesting to compare the Bolshoi results with 
  \citet{Nagai05}, who  present profiles for 8 clusters with almost
the same masses as in the top panel of our
Figure~\ref{fig:SatProfiles}. If we change their normalization for
subhalos selected by present-day circular velocity (their Figure~3) in such a
way that the dark matter profile matches the overdensity of satellites
at the virial radius (we use a factor of 0.8 to match our definition of virial
radius), then there is excellent agreement with Bolshoi, with both
simulations giving the ratio of the dark matter density to the
number-density of satellites $\sim 2$ at $R=0.2R_{\rm vir}$.

\section{Conclusions}
\label{sec:fin}

Using the large halo statistics and high resolution of the Bolshoi
simulation we study numerous properties of halos and subhalos. We
present accurate analytical approximations for such characteristics as
the halo and subhalo abundances and concentrations, the velocity
functions, and the number-density profiles of subhalos. Detailed
discussions of different statistics have already been given in
relevant sections of the text. Here we present a short summary of our
main conclusions.

{\it Velocity function.}  Our main property of halos is their maximum
circular velocity \Vcirc. As compared with virial masses, circular
velocities are better quantities to characterize the physical
parameters of the central regions of dark matter halos. As such, they
are better quantities to relate the dark matter halos and galaxies,
which they host.  We present the halo velocity functions at different
redshifts and show that they can be accurately described by
eqs.~(\ref{eq:nv}-\ref{eq:pars}). The halo circular velocity function
$n(>V)$ declines as a power-law $V^{-3}$ at small velocities and has a
quasi-exponential cutoff at large circular velocities.

{\it Mass function of distinct halos.} We find that the ST
approximation \citep{ST02} gives an accurate fit to the redshift-zero
mass function: errors are less than 10\% for masses in the range
$5\times 10^{10}\Msunh - 5\times 10^{14}\Msunh$. However, the
approximation overpredicts the halo abundance at higher redshifts and
gives a factor of ten more halos than the Bolshoi simulation at
$z=10$. The correction factor eq.~(\ref{eq:STcorrect}) brings the
accuracy of the approximation back to the $\sim 10$\% level for redshifts
$z=0 - 10$. It also breaks the universality of the fit: the mass
function cannot be written as a function of only the rms fluctuation
$\sigma(M)$ on mass scales $M$. These results depend on how halos
  are defined with the Friends-Of-Friends algorithm giving different answers
  than the spherical overdensity method. See Appendix C for details.

  {\it Concentrations of halos.} The halo concentration $c(\Mvir,z)$
  appears to be more complex than previously envisioned. For a given
  redshift $z$, the concentration first declines with increasing
  mass. Then it flattens-out and reaches a minimum of $c_{\rm
    min}\approx 4-5$ with the value of the minimum changing with
  redshift.  At even larger masses $c(\Mvir)$ starts to slightly
  increase. This ``up-turn'' in the concentration is a weak feature:
  the change in concentration is only 20\%.  Moreover, it cannot be
  detected at low redshifts, $z<0.5$. If our estimates are correct, at
  $z=0$ the upturn should start at masses about $\Mvir\sim
  10^{18}\Msunh$ - clusters this massive do not exist. However, at
  $z>2$ the upturn is visible at the very massive tail of the mass
  function.  It is not clear what causes the upturn.  The upturn is
  even stronger for relaxed halos, which indicates that
  non-equilibrium effects cannot be the reason for the upturn.  At
  large redshifts the halos that show the upturn are very rare: their
  mass is much larger than the characteristic mass $M_*$ of halos
  existing at that time. Most of them likely experience a very fast
  growth.  They also represent very high-$\sigma$ peaks of the density
  field.  It is known that the statistics of rare peaks are different
  from those of more ``normal'' peaks \citep{BBKS}. One may speculate
  that this may result in a change in halo concentration. More
  extensive analysis of halo concentrations is given in
  \citet{Prada11}.

  {\it Subhalo abundance.} The cumulative abundance of satellites is a
  power-law with a steep slope: $N(>V)\propto V^{-3}$.  Combing our
  results with those of the Via Lactea-II simulation \citep{VLII}, we
  show that the power-law extends at least from 4~\kms~ to 150~\kms
  for Milky-Way-mass halos and yields the correct abundance of large
  satellites such as the Large Magellanic Cloud \citep{Busha11}. The
  abundance of satellites increases with the circular velocity and
  mass of the host halo as $N\propto V_{\rm host}^{1/2}$. For example,
  this means that in relative units a cluster of galaxies has 2.5
  times more satellites than the Milky Milky Way, in good agreement
  with previous numerical results
  \citep{Gao04}. Eqs.~(\ref{eq:Vsub}-\ref{eq:Vsubacc}) give
  approximations for the abundance of satellites.

{\it Subhalo number-density distribution.} One of the main issues here
is the number-density of satellites relative to the dark matter in the
outer regions of halos. Some previous simulations indicated
substantial (a factor of two) overabundance of satellites around the
virial radius. We do not confirm this conclusion. Our re-analysis of
the Via Lactea-II simulation as well as the results from the Bolshoi simulation
unambiguously show that there is no overabundance of satellites. In
the Via Lactea-II simulation the satellites follow very closely the
distribution of dark matter for radii $R= (0.3-2)\Rvir$. In the
Bolshoi simulation we find a small (10\%) antibias at the virial
radius.

\acknowledgements We thank F.~Prada and A.~Kravtsov for numerous
helpful discussions. We also thank S.~Gottloeber for commenting on our
paper and for allowing us to use his FOF halo catalog.  We are
grateful to P.~Madau and M.~Kuhlen for providing us the halo catalog
of the Via Lactea-II simulation and for discussions. We thank
M.~Boylan-Kolchin for commenting on the abundance of halos in the
Millennium simulations. We thank the reader for reading this long
manuscript.  We acknowledge support of NSF grants to NMSU and NASA and
NSF grants to UCSC.  Our simulations and analysis were done at the
NASA Ames Research Center.

\newpage
\appendix

\section{APPENDIX: Bound Density Maxima halofinder}
We use a parallel (MPI+OpenMP) version of the Bound-Density-Maxima
algorithm to identify halos in Bolshoi \citep{Klypin97}. For detailed
comparison with other halofinders see \citet{Knebe11}. The code
detects both distinct halos and subhalos. The code locates maxima of
density in the distribution of particles, removes unbound particles,
and provides several statistics for halos including virial mass and
radius, as well as maximum circular velocity. The parameters of the
BDM halo finder were set such that the density maxima are not allowed
to be closer than $10\,\kpch$.  We keep only the more massive density
maximum \footnote{We keep the peak that has the largest mass inside a
  sphere of radius $10\,\kpch$} if that happens. This is mostly done
to save computer time. It is also consistent with the force resolution
of Bolshoi. Halo catalogs obtained with a smaller minimum separation
of $7.5\,\kpch$ did not include more halos.

Removal of unbound particles is done iteratively. It goes
in steps:
\begin{enumerate}
\item Find the bulk velocity of a halo: the velocity with which the halo
  moves in space. The rms velocities of individual particles are later
  found relative to this velocity. We use the central region of
  the halo (the 30 particles closest to the halo
  center) to find the bulk velocity.
\item Find the halo radius: the minimum of the virial radius and the
  radius of the declining part of the density profile (radius of the
  density minimum, if it exists).
\item Find the rms velocity of dark matter particles and the circular
  velocity profile. Estimate the halo concentration.
\item Find the escape velocity as a function of radius and remove
  particles that exceed the escape velocity. Use only bound
  particles for the next iteration.
\end{enumerate}
The whole procedure (steps 1--4)  is repeated 4 times. If the mass or radius of a
halo are too small (too few particles), the density maximum is removed
from the list of halo candidates. 

If two halos (a) are separated by less than one virial radius, (b) have
masses that  differ by less than a factor of 1.5,  and (c) have a relative
velocity less than 0.15 of the rms velocity of dark matter particles
inside the halos, we remove the smaller halo and keep only the larger
one. This is done to remove a defect of halo-finding where the same halo
is identified more than once. This removal of ``duplicates'' (halos with
nearly the same mass, position, and velocity) happens only during
major merger events when instead of two merging nearly equal-mass
halos the halo finder sometimes finds 3-5 halos. Unfortunately, this
also has the side effect of removing one of the major merger halos. This is
a relatively rare event and it affects only the very tip of the
subhalo velocity function.

We use the virial mass definition \Mvir~ that follows from the
top-hat model in an expanding Universe with a cosmological
constant. We define the virial radius $R_{\rm vir}$ of  halos as the radius
within which the mean density is the virial overdensity times the mean
universal matter density $\rho_{\rm m}=\Omega_{\rm M}\rho_{\rm crit}$ at that
redshift. Thus, the virial mass is given by
\begin{equation}
    M_{\rm vir} \equiv {{4 \pi} \over 3} \Delta_{\rm vir} \rho_{\rm m} R_{\rm vir}^3 \ .
\end{equation}
Eq.~(\ref{eq:A1}) gives an analytical approximation for $\Delta_{\rm
  vir}$.  For our set of cosmological parameters, at $z=0$ the virial
radius \Rvir~ is defined as the radius of a sphere with overdensity of
360 times the average matter density. The overdensity limit changes with
redshift and asymptotically approaches 178 for high $z$. 

Overall, there are about 10~million halos in Bolshoi (8.8~M at $z=0$,
12.3~M at $z=2$, 4.8~M at $z=5$). Halo catalogs are complete for halos
with $\Vcirc>50$~\kms~ ($\Mvir \approx 1.5\times 10^{10}\Msunh$).  We
do post-processing of identified halos. In particular, for distinct
halos we find their properties (e.g., mass, circular velocity, density
profiles) without removing unbound particles. For most, but not all,
halos it makes little difference. For example, the differences
in circular velocities are less than a percent for halos with and
without unbound particles. Differences in mass can be a few percent
depending on halo mass and on environment.

\begin{figure}[tb!]
\epsscale{0.5}
\plotone{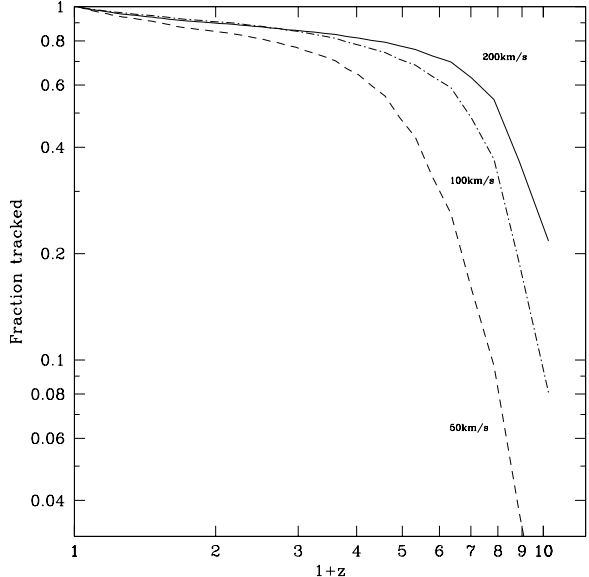}
\caption{Fraction of $z=0$ halos tracked to a given redshift for halos with
  different circular velocities at redshift zero.}
\label{fig:track}
\end{figure}

In order to track the evolution of halos over time, we find and store the
50 most bound particles (fewer, if the halo does not have 50 particles).
Together with other parameters of the halo (coordinates, velocities,
virial mass and circular velocity) the information on most bound
particles is used to identify the same halos at different moments of
time.  The procedure of halo tracking starts at $z=0$ and goes back in
time. The final result is the history ({\it track}) of the major
progenitor of a given halo. The halo track may be lost at some high
redshift when the halo either becomes too small to be detected or the
tracking algorithm fails to find it. A new halo track may be initiated
at some redshift if there is a halo for which there was no track at
previous snapshots (smaller redshifts). This happens when a halo merges
and gets absorbed by another halo. 

With $\sim$180 snapshots stored, the time difference between
consecutive snapshots is rather small. For example, the snapshot
before the $z=0$ snapshot has $z=0.0027$ with a time difference of
37~Myrs.  The difference in time between snapshots stays on nearly the
same level (42-46~Myrs) until $z=0.23$ when it becomes twice as large.
We start with $z=0$ halos and identify them in the previous
snapshot. If a halo is not found at that snapshot, we try the next
one. Altogether, we may try 6 snapshots. Typically, 95\% of halos are
found in the previous snapshot, an additional 2-3\% in the next one and
$\sim$1\% in even earlier ones. Overall, about (0.2-0.3)\% of halos
cannot be tracked at any given snapshot: they are either lost because
their progenitor gets too small or because of numerical problems.  The
number depends on the redshift and on halo
mass. Figure~\ref{fig:track} shows the fraction of halos tracked to
given redshift for halos that exist at $z=0$. More massive halos are
tracked to larger redshifts. Half of all halos with  $\Vcirc=50$~\kms~ are
tracked to $z=4$ and half of all halos with $\Vcirc=200$~\kms~ are tracked to
$z=7$.

\section{APPENDIX: Auxiliary approximations}
For completeness, here we present  some approximations
used in the text.
For the family of flat cosmologies
($\Omega_{\rm M} + \Omega_\Lambda = 1$) an accurate approximation for the
value of the virial overdensity $\Delta_{\rm vir}$ is given by the analytic formula
\citep{BryanNorman98}: 
\begin{equation}
\Delta_{\rm vir} = (18 \pi^2 + 82x -39x^2)/\Omega(z), \label{eq:A1}
\end{equation}
where $\Omega(z)\equiv \rho_{\rm m}(z)/\rho_{\rm crit}$ and $x \equiv
\Omega(z)-1$.

The linear growth-rate function $\delta(a)$ used in $\sigma_8(a)$ is defined as
\begin{equation}
 \delta(a) =D(a)/D(1),
\label{eq:delta}
\end{equation} 
where $a=1/(1+z)$ is the expansion parameter and $D(a)$ is:
\begin{equation}
 D(a) = \frac{5}{2}\left( \frac{\Omega_{\rm M,0}}{\Omega_{\Lambda,0}} \right)^{1/3}
                             \frac{\sqrt{1+x^3}}{x^{3/2}}
                               \int_0^x\frac{x^{3/2}dx}{[1+x^3]^{3/2}},
\end{equation}
\begin{equation}
     x \equiv \left( \frac{\Omega_{\Lambda,0}} {\Omega_{\rm M,0}} \right)^{1/3}a,
\end{equation}
where $\Omega_{\rm M,0}$ and $\Omega_{\Lambda,0}$ are the density
contributions of matter and the cosmological constant at $z=0$ respectively. For
$\Omega_{\rm M}>0.1$ the growth rate factor $D(a)$ can be accurately
 approximated by the following expressions \citep{Lahav91,Carroll92}:
 \begin{equation}
 D(a) = \frac{(5/2)a\Omega_{\rm M}}{\Omega^{4/7}_{\rm M}-\Omega_{\rm \Lambda}
                    +(1+\Omega_{\rm M}/2)(1+\Omega_{\rm \Lambda}/70)},
\end{equation}
\begin{eqnarray}
  \Omega_{\rm M}(a)&=&(1+x^3)^{-1}, \\
\Omega_\Lambda(a)&=&1-\Omega_{\rm M}(a).
\end{eqnarray}
For $\Omega_{\rm M,0}=0.27$ the error of this approximation is less
than $7\times 10^{-4}$.

The Sheth-Tormen approximation \citep{ST02} for the distinct halos
mass function can be written in the following form:
\begin{equation}
   M\frac{dn}{dM} = \Omega_{\rm M,0}\rho_{\rm cr,0}\frac{d\sigma(M)}{\sigma(M)dM}f(\sigma)\qquad
\end{equation}
\begin{equation}
   = 2.75\times 10^{11}(h^{-1}{\rm Mpc})^{-3}\Omega_{\rm M,0}h^2
        \frac{M_\odot d\sigma}{\sigma dM}f(\sigma),
\end{equation}
where $M$ is the halo virial mass  and 
\begin{equation}
   \sigma^2(M) =\frac{\delta^2(a)}{2\pi^2}\int_0^\infty k^2P(k)W^2(k,M)dk, 
\end{equation}
\begin{equation}
   f(\sigma) = A\sqrt{\frac{2b}{\pi}}\left[ 1+(bx^2)^{-0.3} \right]
        x\exp\left({-\frac{bx^2}{2}}\right),
\end{equation}
\begin{equation}
   x\equiv \frac{1.686}{\sigma(M)}, \quad A =0.322, \quad b = 0.707.
\end{equation}
Here $P(k)$ is the power spectrum of perturbations, and $W(k,M)$ is the
Fourier transform of the real-space top-hat filter corresponding to a sphere of 
mass $M$. For the cosmological parameters of the Bolshoi simulation  the rms density
fluctuation $\sigma(M)$ can be approximated by the following expression:
\begin{eqnarray}
\sigma(M) &=& \frac{16.9y^{0.41}}{1+1.102y^{0.20}+6.22y^{0.333}},\quad \\
     y& \equiv& \left[\frac{M}{10^{12}\Msunh} \right]^{-1}. 
\end{eqnarray}
The accuracy of this approximation is better than 2 percent for masses
$M> 10^7\Msunh$.

\section{APPENDIX: FOF and SO masses}
\label{sec:ApB}
\begin{figure}[tb!]
\epsscale{1.1}
\plottwo{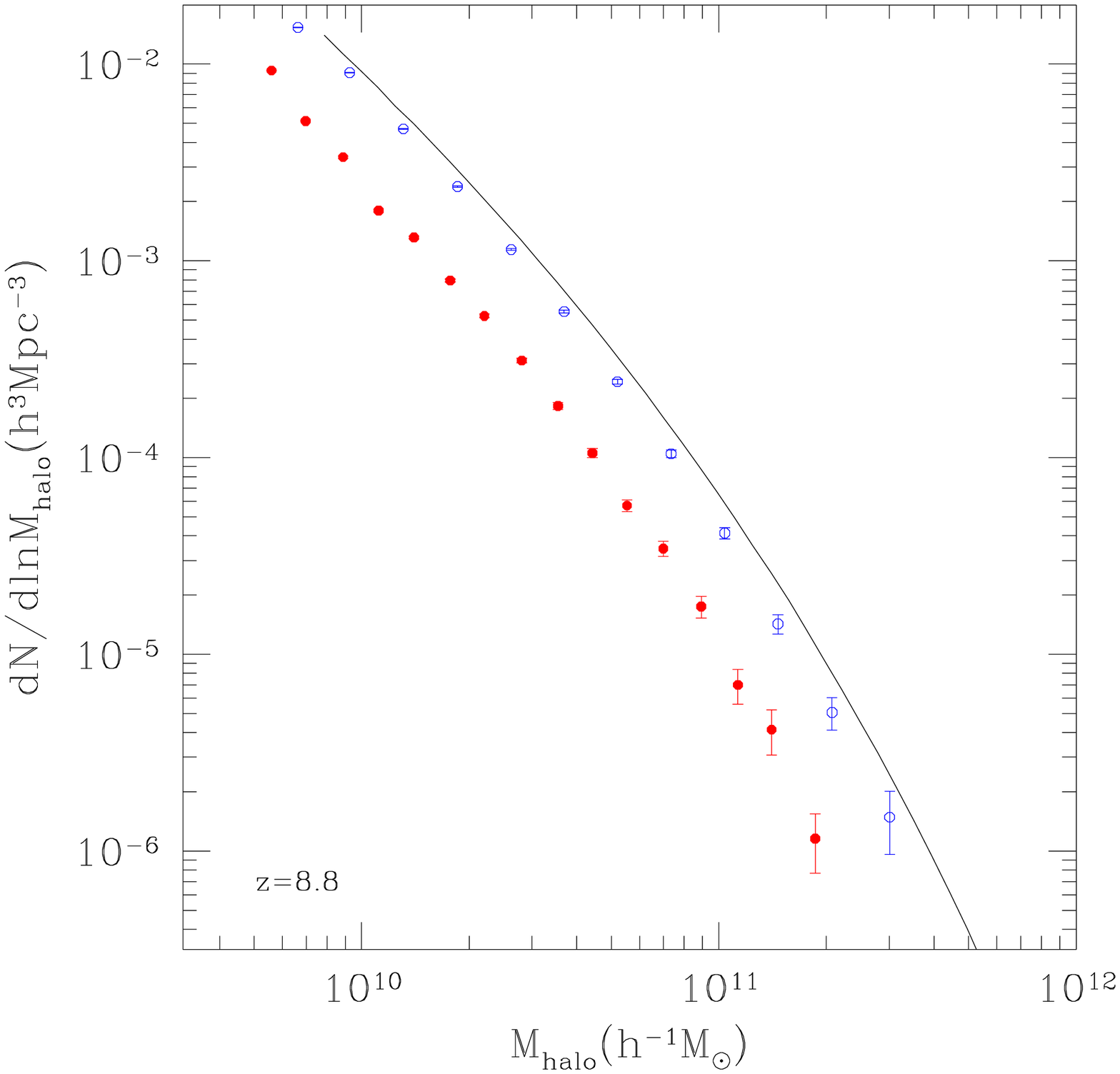}{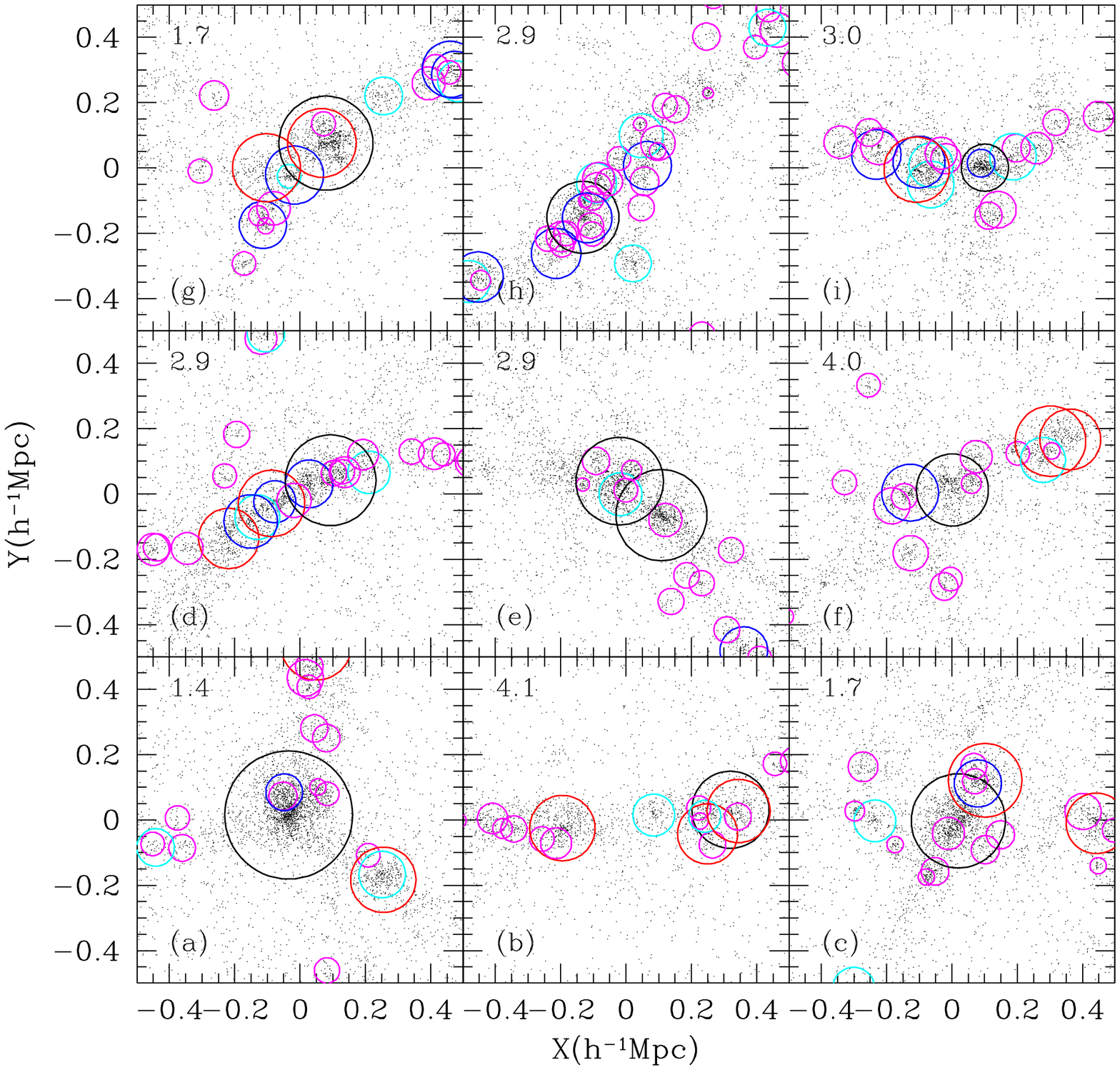}
\caption{{\it Left:} The halo mass function at redshift $z=8.8$. The
  full curve shows the Sheth-Tormen (ST) approximation. Open circles
  show FOF halos identified using the linking length
  $l=0.20$. The spherical overdensity halos are represented by solid
  circles. The ST approximation overpredicts the FOF (SO) masses by a
  factor 1.15 (1.9). {\it Right:} The distribution of mass around 9
  massive halos ($M_{\rm FOF}\approx 10^{11}\Msunh$) at redshift
  $z=8.8$. Each panel shows 1/2 of the dark matter particles in cubes
  of $1\,\Mpch$ size. The center of each cube is the exact position of
  the center of mass of the corresponding FOF halo. The effective
  radius of each FOF halo in the plots is $150-200\kpch$. Circles
  indicate distinct halos and subhalos identified by the spherical
  overdensity algorithm BDM. The radius of each circle is equal to the
  virial radius of the halo. The numbers in the top-left corner of
  each panel show the ratio of FOF mass to that of SO. Panels (a, c,
  g) show relatively good cases when the center of a halo in the
  simulation is close to the center of a FOF-detected halo. Panel
  (e) shows a major-merger: FOF linked the two halos together. In panels (b, d,
  f, h, i) FOF linked together  halos which formed long
  and dense filaments.}
\label{fig:FOFmass}
\end{figure}

In order to clarify the situation with the difference between the
results on the halo mass function in the Bolshoi simulation and in the
ST approximation at high redshifts, we make more detailed analysis of
halos at redshift $z=8.8$.  We also study results obtained using the FOF
halofinder  with three linking-lengths: $l =0.17, 0.20,$ and
0.23. We start by considering only the most massive halos with masses
larger than $10^{11}\,\Msunh$. Each of those halos should have more than
700~particles. The spherical overdensity algorithm (BDM) identified 55
halos above this mass threshold. The FOF found 121, 255, and 602 halos
with linking lengths $l =0.17, 0.20,$ and 0.23 above the same mass
threshold. It is clear that FOF gives significantly higher masses. It
is also very sensitive to the particular choice of the linking length.

At $z=8.8$ the ST approximation predicts a factor of 4 -- 6 more halos as
compared with what we find using the spherical overdensity algorithm.
Because FOF with $l =0.20$ gives about  five times more halos, it makes a
very good match to the ST approximation. This is consistent with
the results of \citet{Cohn08} and \citet{Reed09}. The left panel in
Figure~\ref{fig:FOFmass} shows the mass functions at $z=8.8$. At all masses FOF
with $l=0.20$ is well above the SO results and is  close to the ST predictions.

However, FOF results are very misleading.
We compare the SO halos (as found by the BDM code) with the FOF halos
found using  $l =0.20$. For halos with more than 100 particles both algorithms
find essentially the same distinct halos, but FOF typically assigns
larger masses to the same halos. The right panel in
Figure~\ref{fig:FOFmass} illustrates the point. There is a large
variety of situations. We typically find that when there is a
well-defined halo center  {\it and} the halo dominates its
environment, both the FOF and the SO masses are reasonably consistent
(e.g., panel (a)).  However, FOF has a tendency to link fragments of
long filaments. In such cases the formal center of the FOF halo
may not even be found in a large halo. Surprisingly, there are many
of those long filaments at the high redshifts.

\begin{figure*}[tb!]
\epsscale{1.1}
\plottwo{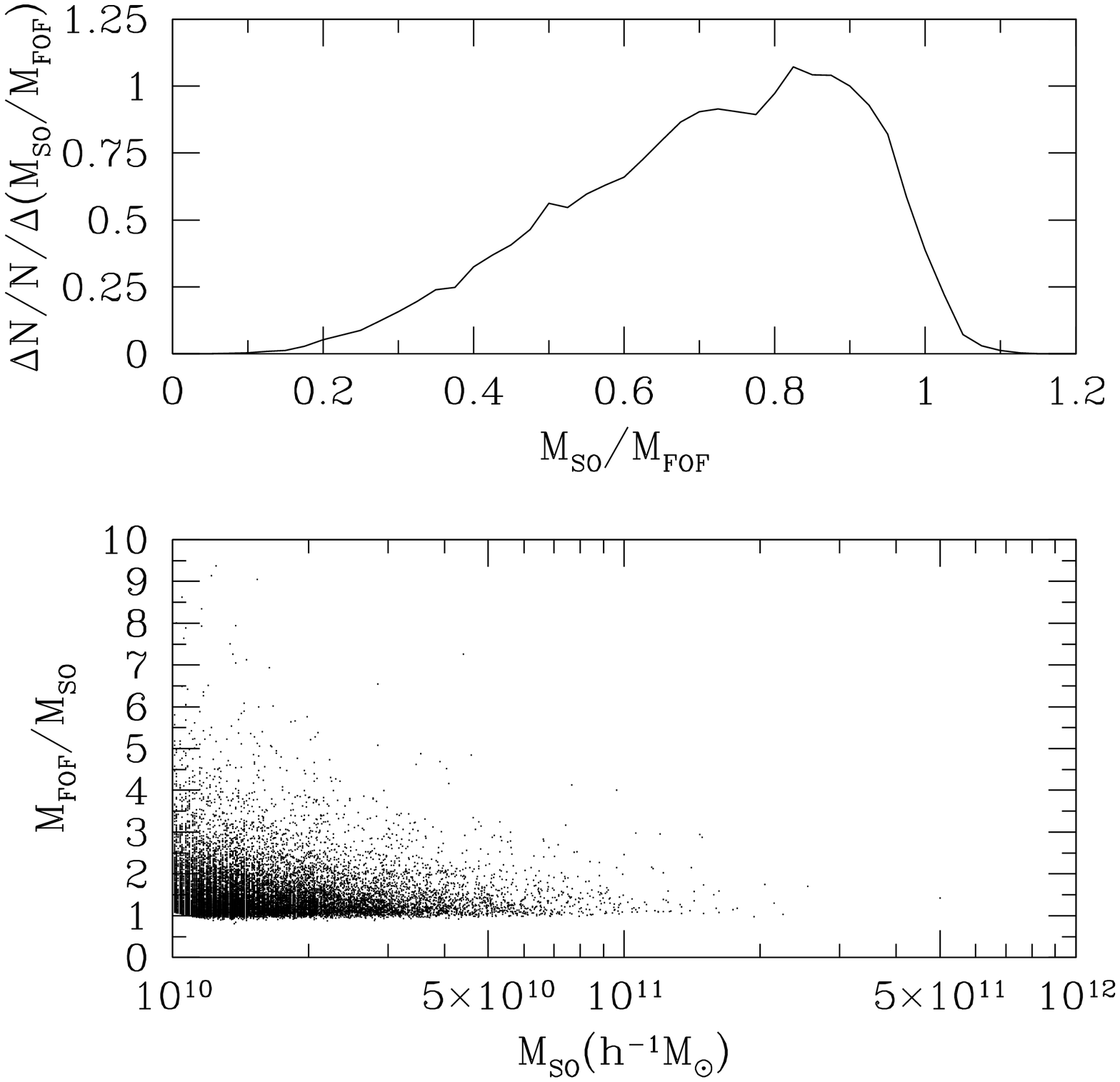}{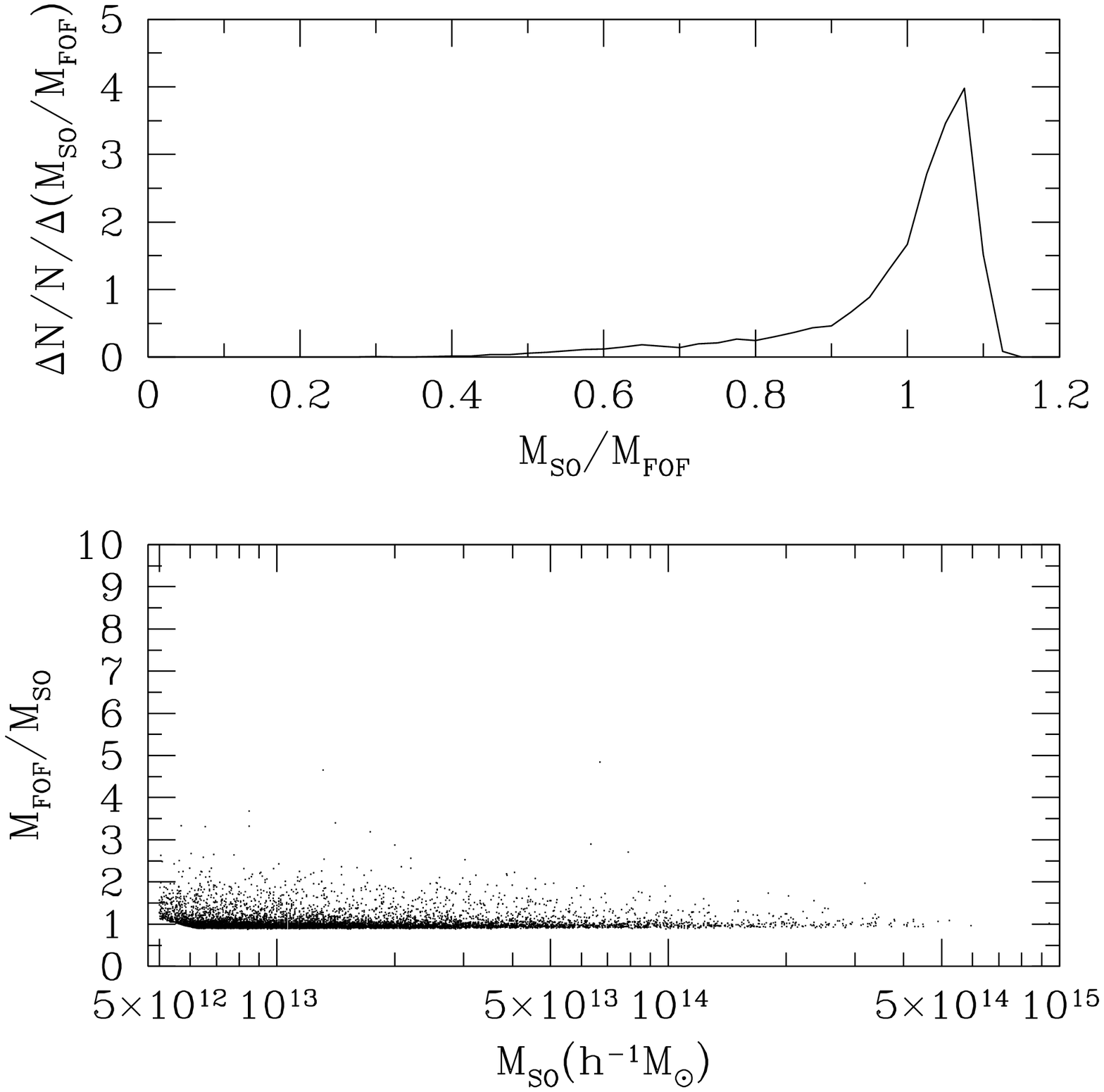}
\caption{{\it Left panels:} Ratio of masses for the same halos
  identified by FOF with $l=0.20$ and by the spherical overdensity BDM
  halofinders at redshift $z=8.8$. On average, the FOF halos have
  masses 1.4 times larger than those obtained using SO. In addition, there is
  a significant spread in the mass ratios. {\it Right panels:} The same
  for $z=0$ with a linking-length $l=0.17$. Top panels show the
  distribution of mass ratios $M_{\rm SO}/M_{\rm FOF}$ with halos
  selected by the SO mass. Bottom panels show the mass ratios for
  individual halos.}
\label{fig:fofbdm}
\end{figure*}

Figure~\ref{fig:fofbdm} presents statistics for the ratios of FOF and SO
masses. Left panels show the most massive 17000 halos with  SO
masses larger than $10^{10}\Msunh$ at $z=8.8$. There is a large spread
of masses and on average FOF masses are 1.4 times larger than the SO
counterparts. We made the same analysis for the most massive 10000 halos with
the SO masses larger than $5\times 10^{12}\Msunh$ at $z=0$ using $l=0.17$ for
FOF. Remarkably, both halofinders produce similar results - a big
contrast with high redshifts.  Overall, there is a small offset in the
mass ratios with SO producing 1.05 times larger masses. However, the
difference is remarkably small.

Thus, as we stated in \S4, FOF halo masses are similar to SO ones at
low redshifts, but systematically larger at high redshifts. 

\end{document}